\documentclass{elsart}
\usepackage{xcolor}
\usepackage{graphicx}
 \usepackage{gensymb}
 
 \newcommand\hl[1]{%
  \bgroup
  \hskip0pt\color{red!80!black}%
  #1%
  \egroup
}

\journal{Computer \& Fluids}

\begin{document}

\begin{frontmatter}

\title{Stratified flow past a sphere at moderate Reynolds numbers}

\author[1]{Francesco Cocetta},
\author[1]{Mike Gillard},
\author[1]{Joanna Szmelter},
\corauth[cor]{Corresponding Author.}\ead{j.szmelter@lboro.ac.uk} 
\author[2]{Piotr K. Smolarkiewicz}

\address[1]{Loughborough University, Leicestershire LE11 3TU, UK}
\address[2]{National Center for Atmospheric Research, Boulder, CO, USA}



\begin{abstract}

A numerical study of stably stratified flows past spheres at Reynolds
numbers $Re=200$ and $Re=300$ is reported. In these flow
regimes, a neutrally stratified laminar flow induces distinctly different
near-wake features. However, the flow behaviour changes significantly
as the stratification increases and suppresses the scale of vertical
displacements of fluid parcels. Computations for a range of Froude
numbers $Fr\in [0.1,\infty]$ show that as Froude number decreases,
the flow patterns for both Reynolds numbers become similar. The
representative simulations of the lee-wave instability at $Fr=0.625$ and
the two-dimensional vortex shedding at $Fr=0.25$ regimes are illustrated
for flows past single and tandem spheres, thereby providing further
insight into the dynamics of stratified flows past bluff bodies. In
particular, the reported study examines the relative influence of
viscosity and stratification on the dividing streamline elevation, wake
structure and flow separation. The solutions of the Navier-Stokes 
equations in the incompressible Boussinesq limit are obtained on
unstructured meshes suitable for simulations involving multiple
bodies. Computations are accomplished using the finite volume,
non-oscillatory forward-in-time (NFT) Multidimensional Positive Definite
Transport Algorithm (MPDATA) based solver. The impact and validity
of the numerical approximations, especially for the cases exhibiting
strong stratification, are also discussed. Qualitative and quantitative
comparisons with available laboratory experiments and prior numerical
studies confirm the validity of the numerical approach.

\end{abstract}


\end{frontmatter}


\section{Introduction}

Stably stratified flows past a sphere display an array of
intricate physical phenomena epitomising flow responses to localised
three-dimensional obstacles. The benchmark emulating a spherical body
moving through a stratified fluid provides a convenient framework for
investigating both the flow behaviour and the numerical methods employed
to simulate flows at a range of Reynolds and Froude numbers. Viscous
stratified flows have important applications in diverse areas of
engineering and geoscience, where examples include wind turbines and
marine appliances submersed in thermoclines \cite{EsmaeilpOE16}, flows
of ocean currents past natural bathymetry \cite{WarnVOM07} as well
as atmospheric flows past mountains and islands \cite{LeeMAP87,Reisner94}. Such
practical interests motivate numerical simulations complementing idealised
experimental studies of flows past a sphere. However, despite a large
body of experimental and numerical work devoted to stratified wakes
behind spheres, the numerical simulations actually resolving the viscous
boundary layer past a sphere are rare due to their technical complexity;
see \cite{vandine18,orr} for recent succinct reviews.

Experimental studies, e.g. \cite{lofquist,honji88,lin92,cho93},
demonstrate the importance of the relative ratio of the characteristic
time scales of the viscous and buoyant forces to the time scale of
inertia. These studies suggest that the structure of a flow depends
largely on the fluid stratification and its interaction with viscous
effects---characterised, respectively, by the Froude $Fr=V_o/Nh$ and
Reynolds $Re=V_o L/\nu$ numbers, with $V_o$, $N$, $h$, $L$, and $\nu$
corresponding to free stream velocity, buoyancy frequency, obstacle
height, characteristic length scale, and kinematic viscosity. As
$Fr\nearrow \infty$, weakly stratified laminar flows demonstrate
three-dimensional structures reminiscent of neutrally stratified flows,
whereas for $Fr\searrow 0$ they become more planar and
two-dimensional vortex shedding can be observed. The latter matches
satellite observations of mesoscale eddies in the lee of mountainous
islands \cite{lee04}. As reported by \cite{lin92,cho93}, a range
of different flow patterns occurs between the two asymptotic flow
regimes. The main responses such as lee waves, obstacle-scale eddies,
and upwind blocking are commonly observed in experiments. However, in
terms of delimiting the transitions between the patterns, experimental
data has been found to be in poor agreement \cite{gus16}.

The pioneering numerical study of a flow past a physical sphere towed
in a linearly density-stratified incompressible fluid at $Re=200$ was
documented by Hanazaki \cite{hanazaki}. Hanazaki's focus was on the spatial
evolution of the near-field flow response to the sphere. Subsequent
investigations spawned a range of distinct numerical strategies
that were driven by diverse interests. Studies focused on temporal
evolution of turbulent wakes formed behind underwater vehicles and
the parameter-dependent far-field wake flow dynamics employed direct
numerical simulation (DNS) \cite{gourlay,brucker,diamessis}, large eddy
simulation (LES) \cite{dommerxx}, and implicit large eddy simulation
(ILES) \cite{diamessis}. These studies played an important role in exploring
turbulent wakes and parameter-dependent flow dynamics for stratified
flows. However, they either completely excluded the sphere from
their simulation domains while relying on suitable initial conditions,
or hybridised the spatial/temporal evolution \cite{vandine18}---in the
spirit of secondary application models where an application of interest
is driven by an output of a different model \cite{smsz09}. Alternative
models which include a sphere immersed in the moving stratified fluid
and utilise high resolution grids required to accurately represent
evolution of boundary layers, separation, and the dynamics of fluids
close to the sphere are computationally demanding; their application
is therefore still relatively uncommon. 


An explicit representation of a sphere’s geometry in a computational
domain can adopt distinctly different approaches based on immersed
boundaries or body-fitted meshes. Seminal DNS studies at $Re=3700$
\cite{pal16,pal17} solved Navier–Stokes equations in cylindrical
coordinates on a staggered grid, with a sphere represented by an
immersed boundary. A broad range of stratifications were
considered with $Fr\in [1,\infty]$ \cite{pal17} and $Fr\in [0.025,
1]$ \cite{pal16}, respectively. These studies provided systematic
investigations of buoyancy effects on the behaviour of wakes. They
confirmed that buoyancy suppresses turbulence as the stratification
increases, however, for low Froude numbers below approximately
$Fr=0.25$, a regeneration of turbulence occurs due to a new regime
of vortex shedding.

Here, we adopt an alternative approach that employs unstructured meshes
fitted to a sphere. It enables simulations fully resolving viscous
boundary layers, important for our particular interest with quantitative 
drag predictions. Similar to our study, recent contributions 
employing body fitted meshes solved incompressible Navier Stokes
equations in the Boussinesq approximation. Among them, simulations
in \cite{orr} were performed for $Fr\in [0.1,\infty]$ at $Re=200$
and $Fr\in [1,\infty]$ at $Re=1000$. These solutions used generalised
curvilinear coordinates and were obtained on a regular non-staggered
grid. A structured spherical grid conforming to the sphere
was also employed in \cite{gus16} with a staggered arrangement for
velocity and pressure.  The solutions in \cite{gus16} used a hybrid
finite-difference discretisation in spherical coordinates, and
investigated flow regimes at $0.005 < Fr < 100$, for $1 < Re < 500$.
Examples of earlier contributions utilising body fitted grids include
\cite{lee04}, reporting simulations of flows at $Re=200$ and $Fr\in
[0.02, 10]$. Their simulations where performed on a structured
mesh generated in a cylindrical domain composed of brick and prismatic
shaped elements with a mixed finite element discretisation for
velocity and pressure. The ancestral work of Hanazaki \cite{hanazaki}
also employed a structured body fitted grid, in a study of flows
at $Re=200$ and $Fr\in [0.25, 200]$. In \cite{hanazaki} the structured
grid was generated by rotating a 2D mesh created with a form of
conformal mapping. Computations used cylindrical coordinates and 
finite difference discretisations of different orders for pressure
and velocity.


The present study builds on our earlier work \cite{szsm19} that documented
density stratified flow calculations for $Re=200$ and for high Reynolds
numbers in a turbulent regime appropriate for ILES. Numerical capabilities
employed here extend the non-oscillatory forward-in-time finite volume
(NFT-FV) approach documented in \cite{szsm19} through MPI parallelisation
of the NFT-FV model. This extension allows for the refinement of earlier
$Re=200$ results, in addition to investigating stratified flow past a
sphere at $Re=300$. Interestingly, for weak stratifications, both cases
have distinctly different wake patterns: where for high Froude numbers
steady state is achieved for $Re=200$, whereas a characteristic structure
of the hairpin vortex shedding is observed for $Re=300$. However, as
stratification increases the flow behaviour of the two cases starts to
show strong similarities. In addition to reducing the time to solution,
the parallelisation benefits the quality of simulation results. Namely,
it allows for the enlargement of the computational domain that reduces the 
impact of the external boundaries. Moreover, it enables the
refinement of the mesh resolution in proximity to the solid boundaries
and in the flow wake region. Last but not least, the
simulations can be run for a longer time; so flow patterns which exhibit themselves
on a longer time-scale can be better detected.

Previous numerical efforts reported in the literature have been conducted
exclusively on structured meshes, whereas the current implementation
permits the use of more flexible unstructured meshes and overcomes the
constraint of the regular grids historically adopted in simulation of
stratified flows. 
Structured meshes (including topologically regular grids) have
merits of simplicity and computational economy for homogeneous flows
in simple domains. However, geophysical flows have a large degree of
heterogeneity, complex geometry of bounding domains, and multiplicity
of scales. In the Earth’s atmosphere and oceans internal gravity
waves are both ubiquitous and intricate, as their occurrence and
form depend on the relative magnitude and structure of ambient
flows/currents, entropy/density stratification, and forcings. The
latter in particular must account for complicated variability of
shorelines and topography/bathymetry. There is also an abundance
of natural multiscale phenomena relevant to weather and climate for
which high variation in mesh resolution is desirable.  Factors
limiting the flexibility of structured meshes in construction of
differential operators are usually related to rigid data structures,
dictated by the topological uniformity of computational cells.
Furthermore, generation of body-fitted structured meshes is restricted
to moderately complex geometries. For geophysical applications,
terrain following coordinates are commonly used to account for
orography or bathymetry. However, this approach cannot be used for
shapes involving overhanging orography, underwater caves and
reservoirs, or submerged objects. The generation of structured body-fitted
meshes for simple geometries such as a sphere is usually accomplished
via conformal mapping. The use of multiblock techniques can add some
flexibility to generate high quality structured meshes for more
complex and multibody configurations, cf. \cite{sz92}, however,
the control of the point distribution, required to capture key flow
features, with conformal meshes is difficult and limited. In contrast,
fully unstructured meshes adopted in this work have flexibility
allowing for easy accommodation of optimal meshes and mesh adaptivity.
Even more importantly, unstructured meshes allow for simulations
involving arbitrary complex geometries.

For the considered flows past a sphere, unstructured meshes
enable local refinement and the optimisation of mesh resolution in
regions where the flow is most complex, such as the boundary layer
and wake.  Additionally, the capability of unstructured meshes to
account for multibody configurations is demonstrated by
extending our study to calculations of stratified flows past two
spheres, for $Re=300$, where mesh refinement in the region between
the two spheres can also be used.

The reminder of the paper is organised as follows. The next
section outlines the governing equations and the numerical approach
employed. Section \ref{sec:Results} documents a systematic numerical study
for a range of Froude numbers $1/Fr \in [0,10]$ for flows past a single
sphere at $Re=200$ and $300$ and two spheres in a tandem configuration
at $Re=300$. Remarks in section \ref{sec:Conclusion} conclude
the paper.

\section{Numerical approach \label{sec:Method}}
\subsection{Governing equations \label{ssec:pdes}}
This study is motivated by atmospheric applications, therefore the actual PDE system
solved in our NFT-FV model is the Lipps-Hemler \cite{LH82,Lipps90} anelastic
system suitable for simulating a variety of mesoscale atmospheric flows 
\cite{smsz11,smszw13,ssx16}. For an ideal atmosphere, the anelastic conservation 
laws of mass, momentum and entropy fluctuations (\ref{anl:mass})-(\ref{anl:th}) 
are compactly written as
\begin{equation} \label{anl:mass}
\nabla \cdot ( {\bf V} \overline{\rho} ) =  0~, 
\end{equation} 
\begin{equation} \label{anl:ped}
{\partial \overline{\rho} V_I \over \partial t} + \nabla\cdot ({\bf V}\overline{\rho}\cdot V_I) =  
 -\overline{\rho}{\partial { \varphi^\prime }\over \partial x_I} +
\overline{\rho} g\frac{\theta^\prime}{\overline{\theta}}\delta_{I3} 
+ (\nabla\cdot{\bf \tau})_I~,
\end{equation}
\begin{equation}\label{anl:th}
{\partial\overline{\rho} \theta^\prime \over \partial t} + \nabla\cdot ({\bf
V}\overline{\rho} \theta^\prime)  =  -\overline{\rho}\bf V \cdot \nabla \theta_e~,
\end{equation} 
where $V_{I\,(=1,2,3)}$ indicate the velocity components in the $x_I$
Cartesian coordinate directions $(x,y,z)$, with time $t$, the Kronecker
delta $\delta_{IJ}$, and potential temperature $\theta$ related to
specific entropy via $ds=c_p ~d\ln\theta$ (with $c_p$ denoting specific
heat at constant pressure). Density and pressure are symbolised
by $\rho$ and $p$, while an overline and subscript ``$e$'' mark the
static reference state and a stably stratified inertial ambient state,
respectively; cf. \cite{smsz11} for a discussion. A density normalised
perturbation pressure in the momentum equation (\ref{anl:ped}) is given
by $\varphi^\prime = (p-p_e)/\overline{\rho}$, while the gravitational
acceleration $g$ enters the buoyancy term on the RHS of (\ref{anl:ped}),
where $\theta^\prime=\theta-\theta_e$ is the potential temperature
perturbation about the ambient profile $\theta_e (x_3)= \theta_o \exp(S
x_3)$;  hereafter stratification $S$ is assumed constant and ``o'' subscript
refers to constant reference values. The deviatoric stress
tensor ${\bf \tau}$ in the last term on the RHS of (\ref{anl:ped})
is defined by
\begin{equation}
\label{tensor1}
{\tau}_{IJ}=\mu\left(\frac{\partial{V}_I}{\partial x_J} + 
\frac{\partial {V}_J}{\partial x_I}\right)~.
\end{equation}
where $\mu=\overline{\rho}\nu$ is the dynamic viscosity. 

The entropy equation (\ref{anl:th}) amounts to the adiabaticity statement
$D\theta/Dt=0$, thanks to the assumption of vanishing thermal
diffusivity---the approximation consistent with laboratory studies
conducted in saline solutions with large Schmidt numbers. The governing
equations are deliberately written in terms of potential temperature
and pressure perturbations to facilitate initial and boundary conditions
as well as to enable a semi-implicit amplitude-error-free representation 
of the buoyant modes. To closely link the anelastic equations
for low Mach number atmospheric flows with studies addressing marine
applications, the governing set (\ref{anl:mass})-(\ref{anl:th}) is
taken in the incompressible Boussinesq limit with $\overline{\rho} =
\rho_o$, $\overline{\theta} = \theta_o$, and $\theta_e(x_3) = \theta_o+S
x_3$. In this limit---valid for displacements of fluid parcels
that are small compared to scale height $S^{-1}$ of the stratified
environment---the equations written in terms of density stratification
or potential temperature stratification are mathematically equivalent.

\subsection{Numerical integration \label{ssec:Numint}}

The governing equations are solved numerically using a semi-implicit
NFT-FV solver, presented in comprehensive detail in \cite{szsm19}. Its key
components consist of the unstructured and hybrid mesh based MPDATA
\cite{smsz05o,smsz05} as the nonlinear advection operator, and a
robust non-symmetric Krylov-subspace elliptic solver \cite{smsz11}
for the Poisson equation for pressure. At the highest level of abstraction,
the NFT model algorithm for the prognostic PDEs (\ref{anl:ped})-(\ref{anl:th})
can be viewed as an analogue to the trapezoidal integral of the Lagrangian 
ODEs underlying (\ref{anl:ped})-(\ref{anl:th}) \cite{smmrg93}. This leads
to a 4x4 implicit linear problem of the form
\begin{equation} \label{linearp}
{\bf\Psi}={\widehat{\bf\Psi}}+0.5\delta{t}{\bf R}({\bf\Psi})~
\end{equation}
in each node of a collocated mesh. In (\ref{linearp}), ${\bf\Psi}$
symbolises the unknown vector of four dependent variables---namely, three
components of {\bf V} and ${\theta^\prime}$---at the future $t+\delta{t}$
time level, ${\widehat{\bf\Psi}}$ marks the explicit part of the solution
including advection of the past ${\bf\Psi}$ values at $t$ combined with
their corresponding RHS forcings, and ${\bf R}({\bf\Psi})$ symbolises
the future RHS forcings comprising the pressure gradient, buoyancy
and the convective derivative of the ambient $\theta_e$.\footnote{The
dissipative forcing of momentum is integrated to $\delta{t}{\mathcal
O}(\delta t)$ and included in ${\widehat{\bf\Psi}}$.} The linear problem
(\ref{linearp}) is inverted algebraically in each node, leading to the
closed form discrete expression
\begin{equation} \label{clfV}
{\bf V}={\widehat{\widehat{\bf V}}}-{\bf C}\nabla \varphi^\prime~,
\end{equation}
where ${\widehat{\widehat{\bf V}}}$ is a modified explicit part
of the solution ${\bf V}$, ${\bf C}$ is a 3x3 matrix of known
coefficients, and $\varphi^\prime$ is the unknown solution for the
pressure variable. Substituting (\ref{clfV}) into the discrete
form of (\ref{anl:mass}) forms the discrete Poisson problem for
$\varphi^\prime$. In the spirit of the exact projection, it ensures
that the updated velocity ${\bf V}$ can satisfy the discrete mass
continuity to a round-off (as opposed to a truncation) error. This is
particularly important for simulation of stratified flows, as it
assures the compatibility of conservative MPDATA advection with the mass
continuity and eliminates fictitious buoyancy forces in regions of locally
undisturbed isentropes. The updated velocity is subject to Dirichlet
boundary conditions, which implies consistent Neumann boundary conditions
on the resulting pressure field. The reader interested in further details
of semi-implicit NFT integrators and the associated elliptic solvers is
referred to \cite{szsm19,szsm11,smgrmr97} and references therein.

Spatial discretisation adopts a median-dual finite-volume approach with
edge-based connectivity outlined in \cite{szzhsm15}. The primary meshes
employed in this work, illustrated in Fig.~\ref{mesh}, combine prismatic
elements near solid boundaries and tetrahedral elements elsewhere. The
median-dual cell faces are constructed from the primary meshes, where
the midpoint of each edge is connected both to the barycentres of the
surrounding polyhedra and to the centres of the polygonal faces sharing
the edge. Noteworthy is the synergism of the collocated arrangement of
all dependent flow variables and the proven ILES property of MPDATA.
On one hand, the data collocations enables algebraic inversion of the
linear problem (\ref{linearp}) at the heart of the semi-implicit NFT
integrators, which generally can be quite complex \cite{PSW08,SKW19}.
On the other hand, collocated meshes admit non-trivial null spaces of the
discrete differential operators, including in particular the generalised
Laplacian that forms the viscous terms. The latter can manifest as a
noise at fine scales near the limit of mesh resolution, even for low
Reynolds number flows that are  otherwise well resolved. The solution
regularisation at the limit of mesh resolution characteristic of ILES
remedies this problem. Importantly, as the MPDATA ILES regularisation
is adaptive (as opposed to additive) to explicit dissipation, it has
a negligible impact on the resolved scales \cite{MSS99,MSW06,Domar03}. In
effect, our NFT-FV schemes benefit from the collocated data arrangement
while being inherently stable even with low-numerical-diffusion
\cite{ssx16,szsm19,kuhn19}.

While NFT codes operating on Cartesian meshes and exploiting the
advantages of the distributed memory paradigm are well established,
the successful implementations of NFT-FV codes for unstructured meshes
on multi-node architectures are more recent. To date these efforts have
focused on global models for the Earth's atmosphere with meshes that
are quasi-uniform in the horizontal and prismatic in the vertical
\cite{smetal16,kuhn19}. In this paper, we complement our earlier
work with new NFT-FV model developments that provide capabilities
for parallel simulations on fully unstructured, irregular meshes. We
utilise the multilevel graph partitioning scheme supplied by the MeTis
package \cite{metis99} to assign partitions of the computational spatial
domain. This library effectively balances the sub-domain workloads
for irregular meshes, while minimising the number of edges crossing
partition boundaries.

The NFT-FV solver utilises a localised stencil on collocated data, and
an auxiliary bespoke OpenMP parallelised code configures partition
halos, with a halo depth of one. The second order operators found
in (\ref{anl:mass}-\ref{anl:th}) and the Poisson equation therefore
require an internal halo exchange after applying first order operators,
but benefit from a reduced memory footprint of internal variables.
Halos operate on a partition-to-partition exchange basis and halo points
are chunked in the data-structure. Most MPI communications exchange
information only between immediate neighbours, with some limited global
communications as required by the Krylov-subspace solver---all of which 
utilise the standard MPI library.

\subsection{Problem formulation}  \label{ssec:pformul}

Simulation results for neutrally and stably stratified flows past a single 
sphere are discussed first, due to its extensive existing research history and the availability of pertinent
data. Following \cite{johnson1999flow,tomboulides00}, the cuboidal 
computational domain has dimensions $[-15,25]\times[-15,15]\times[-15,15]$
in units of a solid sphere diameter $D$ (set to unity) placed at the origin.
Within a $0.2D$ thick region from the sphere surface, the hybrid primary
mesh is composed of 10 layers of prismatic elements with triangular
bases. Everywhere outside this region, the mesh is composed of tetrahedral 
elements. The mesh contains $1915318$  nodes, with decreasing inter-nodal
distance in the proximity of the sphere surface and the flow wake, with the
shortest edge length of $0.005D$. No-slip boundary conditions are applied on
the sphere's surface and free-slip conditions on the spanwise faces of
the domain. A constant free stream velocity $V_e=V_o=(1,0,0)\ ms^{-1}$
is prescribed at the streamwise boundaries of the domain, while absorbers
applied at a distance of $2.0D$ from inlet and outlet boundaries attenuate
the solution toward the free stream condition. The potential flow solution
initialises all presented simulations.

In the quantitative analysis, the drag $C_d$, side force $C_s$
and lift $C_l$ coefficients, together with the Strouhal number $St$,
are used to evaluate the results. For a flow impinging on a sphere,
the force coefficients are defined as
\begin{equation} 
\label{drag_coeff}
C_\zeta=\frac{F_\kappa}{0.5\rho_o V_o^{2}A}~,
\end{equation}
where $A=\pi (D/2)^2$ is the reference area, and $\zeta$ indicates indices $d$,
$s$ or $l$. The forces $F_\kappa$---with $\kappa=1,2,3$ corresponding to drag, side and lift
forces acting on the sphere---are computed according to
\begin{equation} 
\label{drag_forces} 
F_\kappa=-\oint \rho_o\varphi\, n_\kappa dS \quad \textrm 
+\oint\mu\left( \frac{\partial V_\kappa}{\partial x_J} 
+ \frac{\partial V_J}{\partial x_\kappa}\right)n_J dS~,
\end{equation}
where for any index $\iota$, $n_\iota$ is the $\iota^{\rm th}$ component of the
unit vector normal to the surface of the face $dS$ of the finite-volume
element and repeating indices imply the summation.

The Strouhal number is defined as $St=f D/V_o$, where $f$ is the
frequency of the vortex shedding. Since $V_o$ and $D$ are equal to one,
the Strouhal number can be directly derived from the frequency spectrum
of the relevant coefficient. We also note that for the sphere the Froude
and Reynolds numbers become $Fr= 2V_o/ND$ and $Re=V_o D/\nu$ respectively.

\section{Results} \label{sec:Results}

\subsection{Flow past a sphere at $Re=200$}
The steady state behaviour of a neutrally stratified flow past a
sphere at the Reynolds numbers between $20$ and $210$ is characterised
by the axisymmetric flow in the wake region. It is well studied by
means of direct numerical simulations \cite{bagchi} and verifiable by
experimental observations \cite{taneda1956experimental}. Computations
with the NFT-FV scheme for $Re=200$ and $Fr=\infty$ are documented in
\cite{szsm19}. They show that the resulting drag coefficient $C_d=0.774$,
the length $L_r=1.42D$ of the steady axisymmetric recirculation bubble
behind the sphere, and the boundary layer separation angle $\phi_s=116.6\degree$
are in agreement with values obtained from the experiment \cite{Nakamura}
and other numerical studies, e.g.  \cite{fornberg88,tomboulides00,bagchi};
see section 4.2 in \cite{szsm19} for further comparisons and references.

In line with \cite{lin92}, the NFT-FV computations indicate that
for Froude numbers higher then  $Fr \sim 10$, the flow resembles the
neutrally stratified case; and, as the Froude number decreases
the density stratification starts to significantly impact the flow
behaviour \cite{lofquist,lin92,cho93}. For $Re=200$, three dominant
flow patterns have been identified in \cite{lin92} throughout the
Froude number regimes. However, the transitions between these flow
patterns are not clearly defined and exhibit intermediate flow features
leading to alternative classifications. For example, \cite{gus16}
distinguishes further different flow characteristics for low and
moderate Reynolds numbers. Our simulations for a range of $Fr$ were
carried out using the NFT-FV scheme until flow patterns have been
established. The non-dimensional simulated time of $T=t\,V_o/D=30$
was used in \cite{szsm19}. The parallel implementation of the scheme
enabled fivefold $T=150$ to tenfold $T=300$ extensions of the simulated
time, used in this work.
\bigskip

\begin{figure}[h]
\centering
\includegraphics[width=0.99\linewidth]{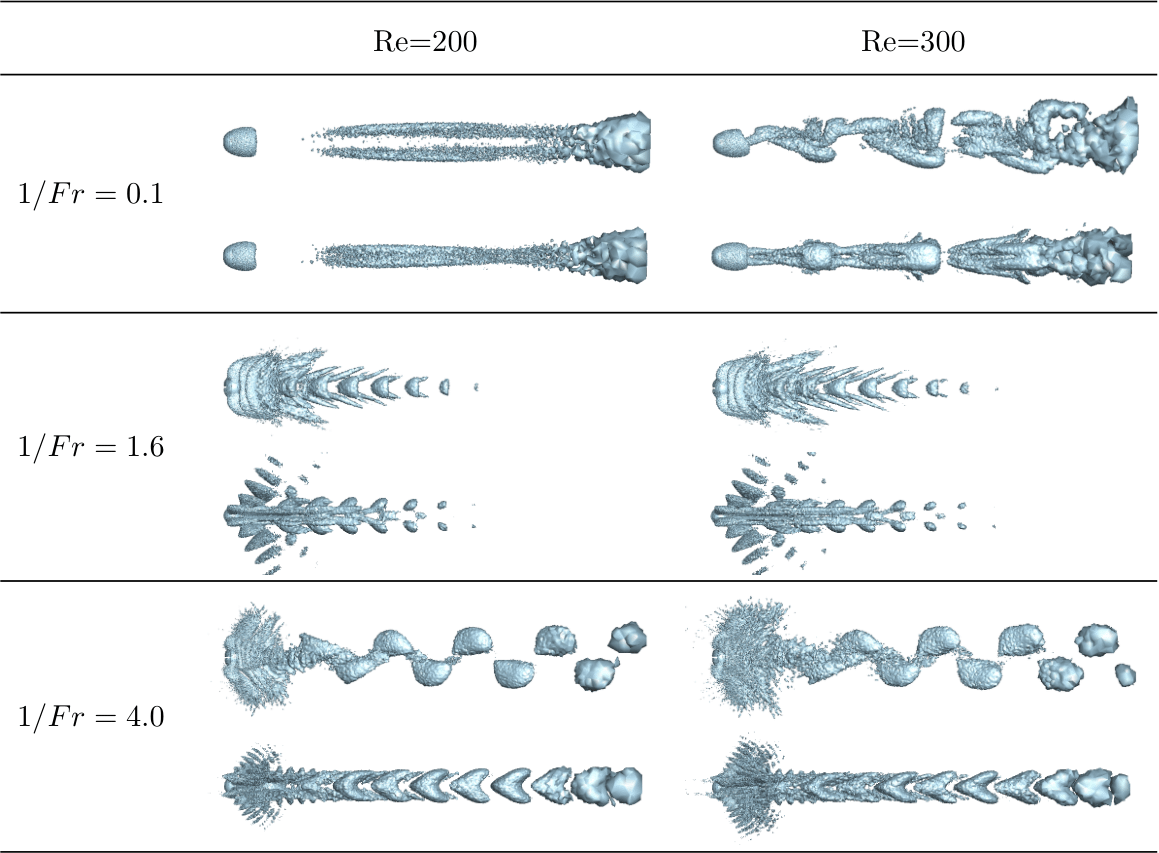}
\caption{A visualisation of the wake vortex structures at $Re=200$ (left)
and $Re=300$ (right). For each case the upper and lower images illustrate 
the flow in the horizontal and vertical central planes, respectively.
\label{wakes} }
\end{figure}

Numerical simulations of stably stratified flows past a sphere for moderate Reynolds numbers, 
especially for $Re=200$, have been reported before (e.g.\ \cite{orr,gus16,pal16,pal17,lee04,hanazaki}) but, to the best of our knowledge, this study together with our earlier investigation in 
\cite{szsm19} are the first to document such simulations using unstructured meshes. 
The results presented here have been carefully selected to enable their verification
against the existing experimental and numerical data and to provide a systematic illustration
of the quality of solutions achievable on unstructured meshes, for the distinct flow regimes 
occurring at $Fr\in [0.1,\infty]$. Especially noteworthy features covered in the following examples are:
cases of symmetrical flows obtained on fully irregular meshes, the ability
of unstructured meshes to capture detailed flow features, and an excellent accuracy of 
the computed drag.

\begin{figure}
\centering
{\includegraphics[width=0.99\linewidth]{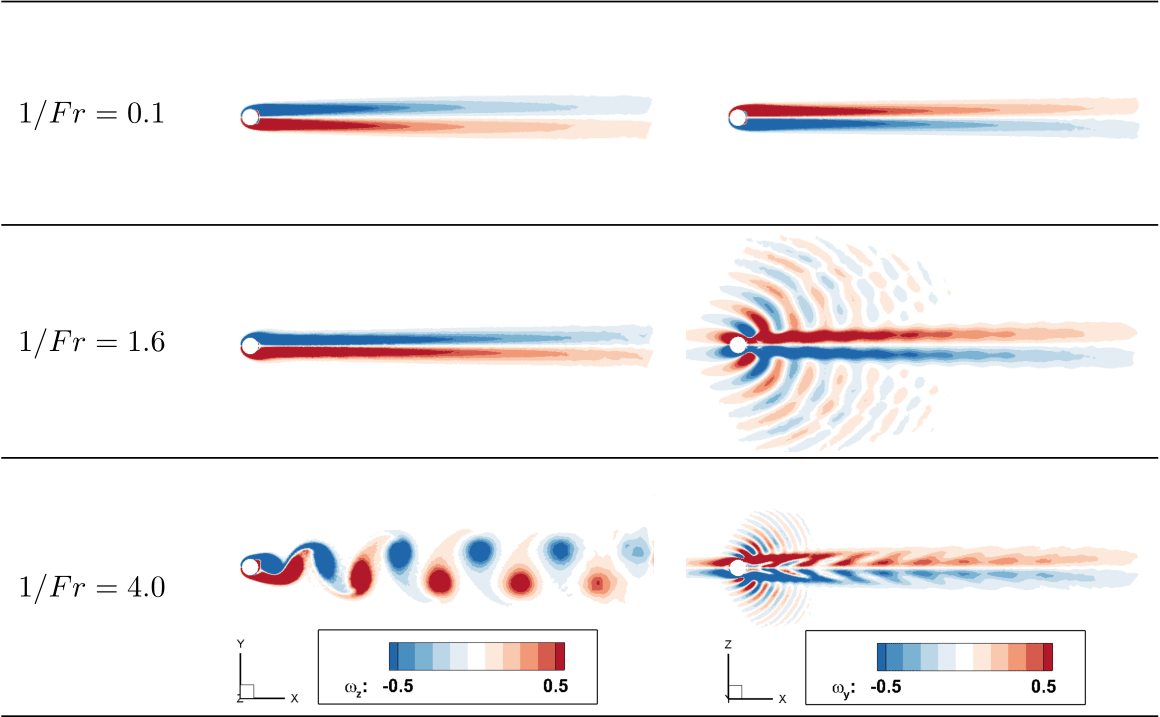}
\caption{Contours of instantaneous vorticity (components normal
to the figure plane) of the wake vortex structures at $Re=200$ 
for $1/Fr=0.1, 1.6, 4$ in central horizontal and vertical plane on
the left and right, respectively. \label{vorticity} }}
\end{figure}

Figure~\ref{wakes} shows an instantaneous level set of
the positive second invariant of the deformation tensor;
$Q=0.5(\Omega_{IJ}\Omega_{IJ} - {\mathcal D}_{IJ} {\mathcal
D}_{IJ})$, where ${\mathcal D}_{IJ}$  is half of the tensorial part
of (\ref{tensor1}), and $\Omega_{IJ}=0.5(\partial{V_I}/\partial{x_J}
-\partial{V_J}/\partial{x_I})$ are entries of the rotation tensor.
This presentation is known as the $Q$-method for vortex identification
\cite{jeong1995identification}. The left column in Fig.~\ref{wakes}
illustrates the 3D structures of the wake flow obtained for $Re=200$
in horizontal and vertical planes at the reciprocal Froude numbers
$1/Fr=\{0.1, 1.6, 4\}$,\footnote{Consider that $1/Fr$ can be thought
of as a normalised dominant wave number.} representative of the
three flow patterns identified in \cite{lin92}. Figure~\ref{vorticity}
illustrates the corresponding vorticity patterns. For the consecutively
increasing values of $1/Fr$ the three patterns represent, respectively, a
non-axisymmetric attached vortex, lee-wave instability, and two-dimensional
vortex shedding. They are in agreement with the experimentally-based 
schematics provided by Figs. 2 and 3 in \cite{lin92}.

\begin{figure}
\centering
\includegraphics[width=0.39\linewidth]{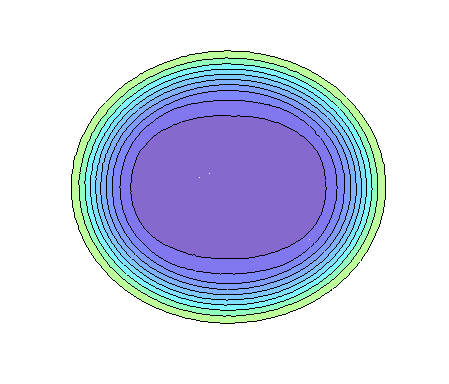}
\includegraphics[width=0.39\linewidth]{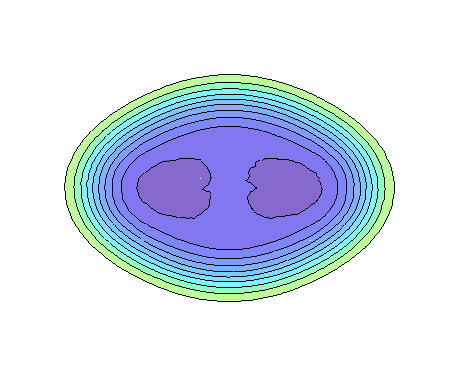}
\includegraphics[width=0.12\linewidth,height=0.38\linewidth]{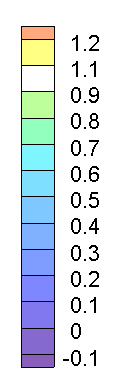}
\caption{Contours of streamwise velocity component in the $x=0.51$ 
plane; $Re=200$, $1/Fr=0.4$ (left) and $1/Fr=0.8$ (right) with respective
(horizontal, vertical) axes equal $(1.42, 1.23)$ and $(1.48, 1.02)$.\label{dou04}}
\end{figure}

Due to the buoyancy-induced gravity waves radiating in the lee, even
for weak stratification $1/Fr=0.1$, the flow past a sphere loses
its axisymmetric character even though viscous effects are still
strong. Planar symmetry of the wake can be observed separately in
the horizontal and vertical central planes in both the 3D images and
vorticity plots, representing a steady recirculation bubble formed behind
the sphere. Noteworthy, the symmetry of the solution is not affected
by the use of irregular unstructured meshes. As the Froude number
decreases, the separation point on the vertical plane moves towards
the rear of the sphere \cite{lin92} and the breaking of axial symmetry
becomes more pronounced. In particular, at $1/Fr=0.4$ the vertical
deformation results in a characteristic doughnut-shaped attached vortex,
illustrated in the streamwise velocity plot at the $x=0.51$ plane in the 
left image of Fig.~\ref{dou04}. At $1/Fr=0.8$ the increased stratification
causes further deformation, such that the doughnut-shaped vortex becomes
doubly connected, leaving two attached vortices as shown in the right
image of Fig.~\ref{dou04}. These results confirm the three-dimensional
interpretative experimentally-based sketches in Figs.~20a,b in \cite{lin92}. 
Similarly to the experimental observation (Fig.~19) and corresponding  
sketch (Fig.~20c) in \cite{lin92}, at $1/Fr=0.8$ the NFT-FV computations 
indicate the formation of the bulge-like particle streak pattern (not shown).

\begin{figure}
\centering
\includegraphics[width=0.69\linewidth]{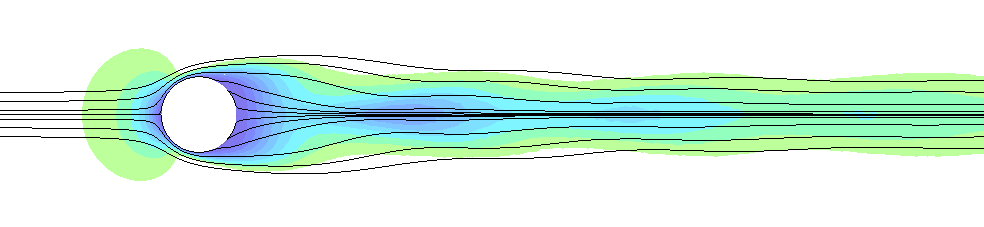}
\includegraphics[width=0.69\linewidth]{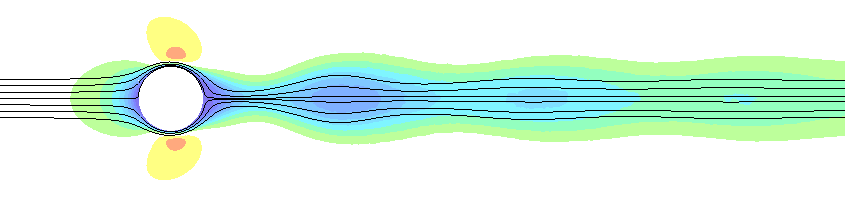}
\includegraphics[width=0.59\linewidth]{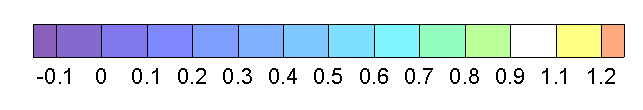}
\caption{Contours of streamwise velocity component and the associated 
streamlines plotted in the planes $z=0$ (top) and $y=0$ (bottom), 
respectively; $Re=200$ and $1/Fr=1$. Streamwise direction extent is
$x\in [-2.7, 10.4]$ (the same in all akin figures). \label{horivelo1a}}
\end{figure}

The bulge-like particle streak pattern also occurs at $1/Fr=1$, when
the lee-wave amplitude approaches its maximum. Figure~\ref{horivelo1a}
presents the streamwise velocity plots and the associated streamlines
at the planes $z=0$ and $y=0$. The bulge-like particle streak pattern
can be clearly seen in the lower panel and the streamlines show that
the flow remains attached to the sphere, while the separation angle
moves towards the aft stagnation point. The patterns represent a
transition between the non-axisymmetric and the lee waves instability
regimes. They are in good agreement with those observed experimentally;
cf. Fig.~18 ($Re=859$, $1/Fr=1.16$) in \cite{lin92}, Fig.~1b ($Re=287$,
$1/Fr=1.136$) in \cite{honji88}, and Fig.~6iv ($Re=329$, $1/Fr=1.11$)
in \cite{cho93}. The flow patterns in the horizontal and vertical
planes are now distinctly different. The special case of $Fr=1$
is further discussed in \cite{szsm19} in the context of the dividing
streamline \cite{HS80} and the linear theories. Consistently with results reported
in \cite{cho93}, at $1/Fr\approx 1.2$, the wake collapses resulting in a
``bow-tie''-shaped separation line, with a bulge at a distance of $1.5D$
from the rear stagnation point.

\begin{figure}[h]
\centering
\includegraphics[width=0.69\linewidth]{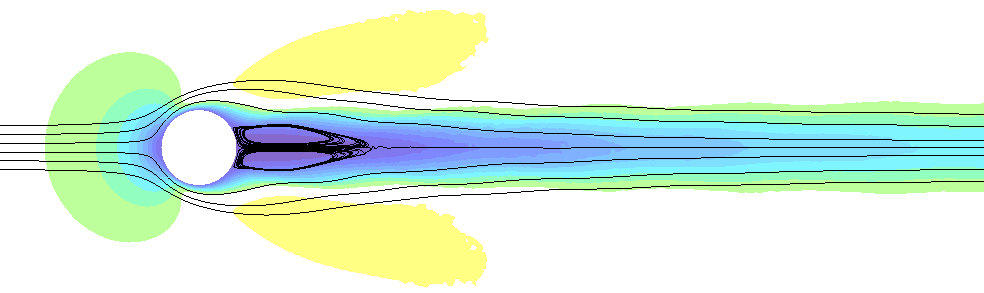}
\includegraphics[width=0.69\linewidth]{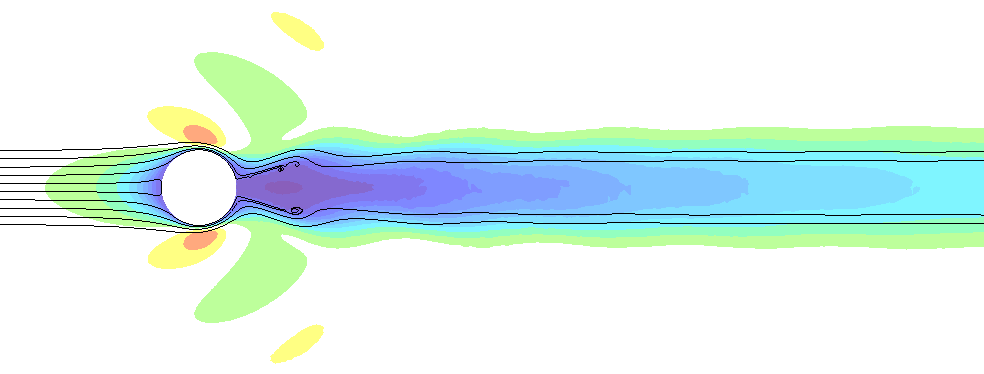}
\includegraphics[width=0.59\linewidth]{figures/horizvelo1lab.png}
\caption{Contours of streamwise velocity component and the associated 
streamlines plotted in the planes $z=0$ (top) and $y=0$ (bottom), 
respectively; $Re=200$ and $1/Fr=1.6$.\label{horivelo16a} }
\end{figure}

The propagation of the lee waves becomes profound between $1/Fr=1.6$
and $1/Fr=2$. The 3D plots in Fig.~\ref{wakes}, vorticity patterns
in Fig.~\ref{vorticity}, and the streamwise velocity components plots
with associated streamlines in Fig.~\ref{horivelo16a} were obtained for
$1/Fr=1.6$. They are representative of the lee-wave instability regime
and show similar features to experimental patterns in Fig.~14 ($Re=500$,
$1/Fr=2.63$) in \cite{lin92}, Figs.~6ii,iii ($Re=329$, $1/Fr=1.66$) in
\cite{cho93}, and Fig.~1a ($Re=144$, $1/Fr=2.27$) in \cite{honji88}. Here,
the concept of the dividing streamline becomes relevant to the resulting
flow pattern. The fluid parcels above the height of the dividing
streamline have sufficient kinetic energy to flow over the top of the
sphere and, due to the presence of lee waves, have a down-slope velocity
greater than the mean stream; cf. \cite{cho93}. The attached recirculation
becomes elongated, Fig.~\ref{horivelo16a}, while the lee waves in the
vertical centre-plane are clearly visible, although dissipated further
downstream. The characteristic overturning motion of fluid parcels,
due to the combined effects of buoyancy and shear \cite{cho93,lin92},
are represented by tiny recirculation regions that can be
seen in the lower image of Fig.~\ref{horivelo16a}. They give the name        
to the ``lee-wave instability regime" \cite{lin92}.\footnote{Note that the
streamline plots are constructed from data stored on irregular meshes,
which affects their symmetry.} 

\begin{figure}[h]
\centering
\includegraphics[width=0.8\linewidth]{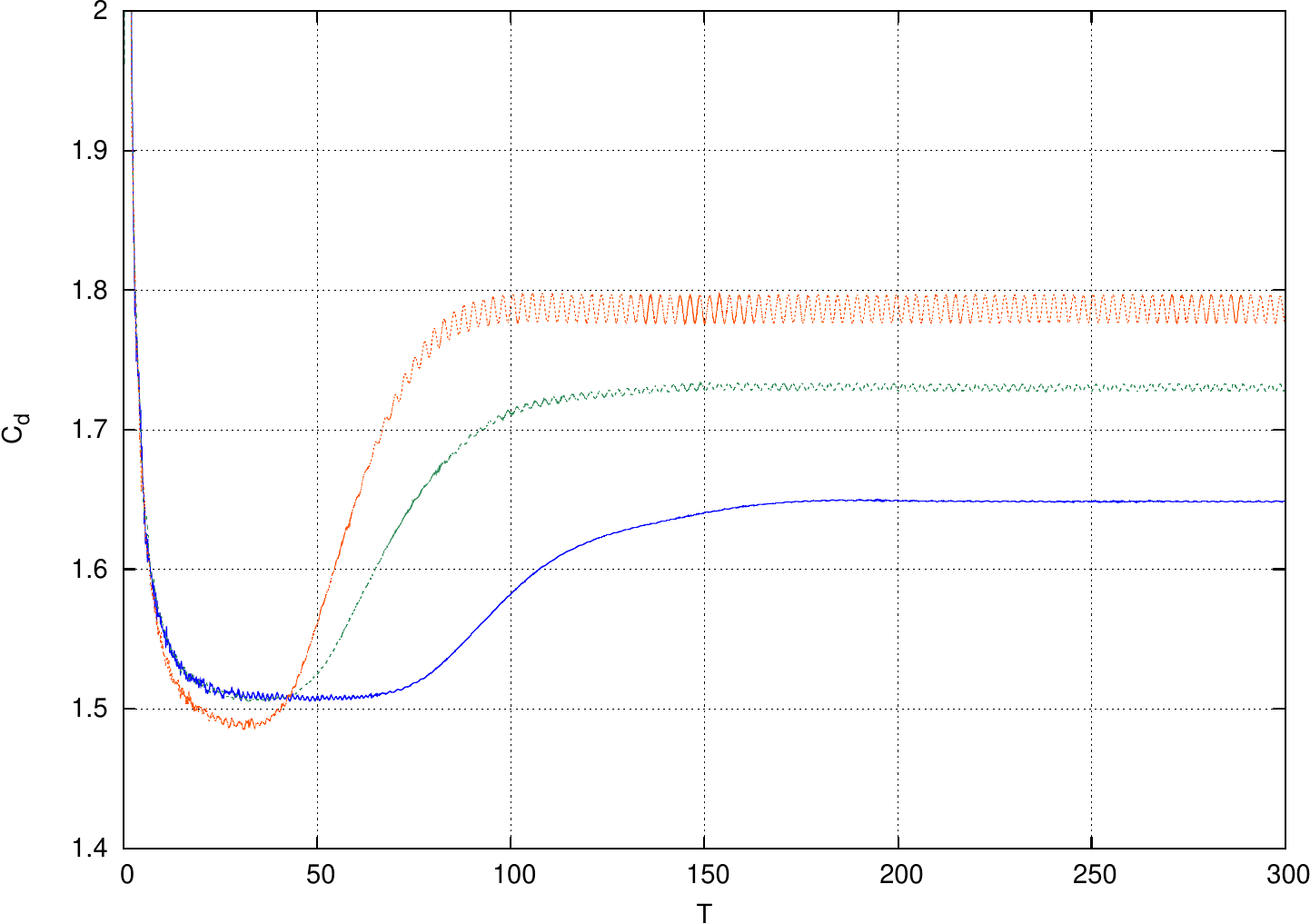}
\caption{Drag coefficient, $C_d$ in (\ref{drag_coeff}), history for 
$Re=200$ and $1/Fr=2.25, 2.5, 3$ respectively marked with solid 
(blue), dashed (green) and dotted (red) lines. \label{cdhistory200}}
\end{figure}

\begin{figure}[h]
\centering
\includegraphics[width=0.69\linewidth]{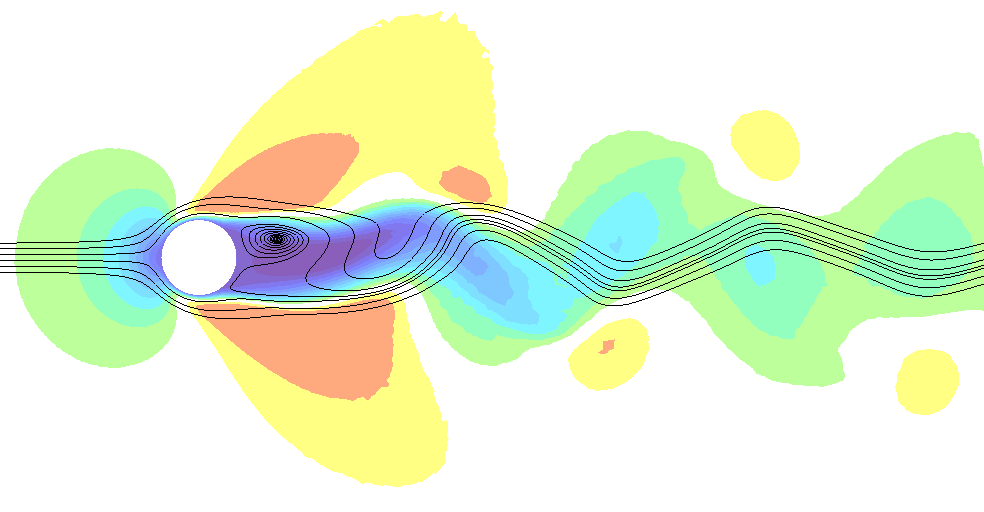}
\includegraphics[width=0.69\linewidth]{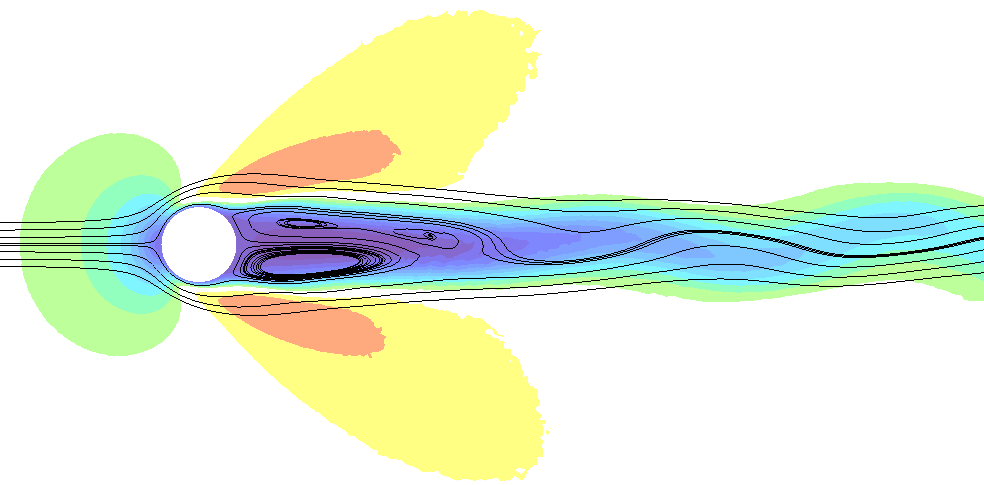}
\includegraphics[width=0.59\linewidth]{figures/horizvelo1lab.png}
\caption{Streamwise velocity component and the associated streamlines in the $y=0$ 
plane; $Re=200$, $1/Fr=2.25$ (top) and $1/Fr=2$ (bottom). \label{horivelo25a}}
\end{figure}

Further increasing the stratification results in a predominantly
two-dimensional flow pattern in the vicinity of the $z=0$ plane, with vertical
motions that become weaker as the stratification increases. Moreover,
computations indicate that starting from about $1/Fr=2.25$, the flow
looses the steadiness characteristic of the neutrally stratified flow
at $Re=200$. The influence of selected stratification levels on the
drag coefficient $C_d$ is illustrated in Fig.~\ref{cdhistory200} for
computations simulating the non-dimensional time $T=300$. The figure
shows that the time from the model initialisation to the onset of 
flow periodicity depends on the stratification. For $1/Fr=2.25$
the periodicity of the flow is hardly visible, as the amplitude in the
$C_d$ plot is almost one order smaller than for $1/Fr=2.5$. However,
examination of the corresponding streamlines and streamwise velocity
at the central horizontal plane, Fig.~\ref{horivelo25a} top, reveals
that the periodic two-dimensional vortex shedding already takes place at
$1/Fr=2.25$, while for $1/Fr=2$, Fig.~\ref{horivelo25a} bottom, the flow 
is still steady but its symmetry in the horizontal plane has been already 
broken.

The NFT-FV results confirm the experimental \cite{lin92} and numerical
\cite{gushchin06} findings reporting that the flow remains stationary
until $1/Fr \approx 2.5$, while computations in \cite{lee04} indicate
that the flow remains stationary for $1/Fr<2$. At $1/Fr=4$, illustrated
in Figs.~\ref{wakes} and \ref{vorticity}, the flow follows a distinct
two-dimensional vortex shedding pattern in the horizontal central
plane. The fluid parcels in the central part of the flow, between the
dividing streamlines, do not have sufficient kinetic energy to travel over
and under the sphere. Strong stratification forces them to flow around,
resulting in a motion similar to a flow around a cylinder. In contrast,
the parcels travelling above and below the dividing streamlines appreciate
the obstacle as a hill of a moderate height which gives rise to the
vertical pattern of linear gravity waves characteristic of flow past
isolated mountains. Both horizontal and vertical patterns are clearly
visible in Figs.~\ref{vorticity} and \ref{wakes}. For $1/Fr=4$ the
vorticity plots illustrate a significant decrease of the gravity waves'
amplitudes and lengths by comparison with results obtained for $1/Fr=1.6$. 
Similar patterns to those shown in Fig.~\ref{wakes} for this flow regime
were obtained both computationally (Figs.~5c-d in \cite{gus16}) and
experimentally (Fig.~10e in \cite{lin92}, and Fig.~5a in \cite{cho93}
with $Re=1961$).

\begin{figure}[h]
\centering
\includegraphics[width=0.48\linewidth,height=0.17\linewidth]{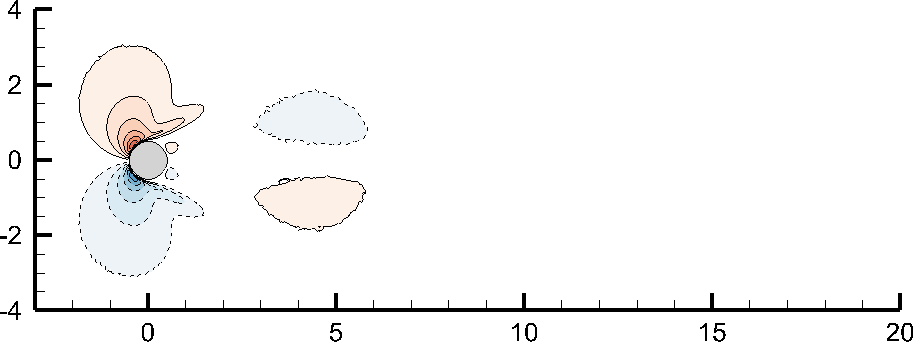}
\includegraphics[width=0.48\linewidth,height=0.17\linewidth]{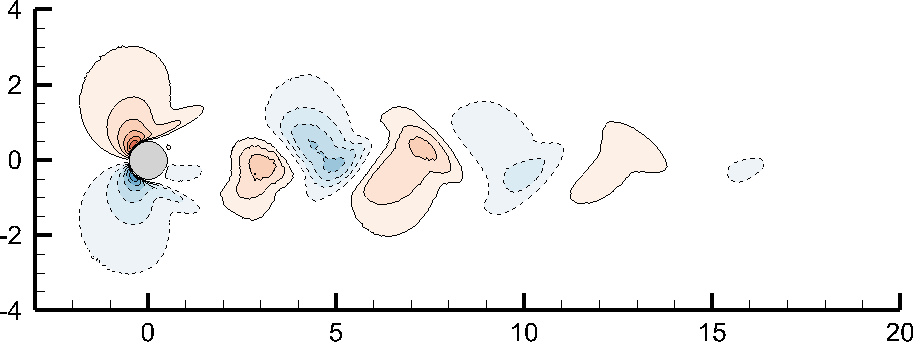}
\includegraphics[width=0.48\linewidth,height=0.17\linewidth]{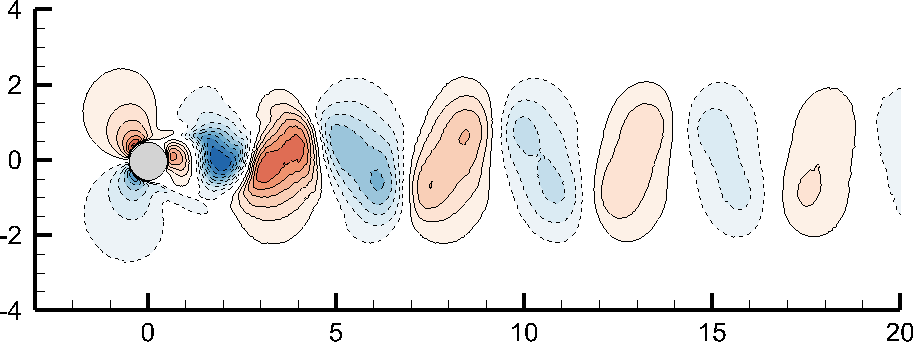}
\includegraphics[width=0.48\linewidth,height=0.17\linewidth]{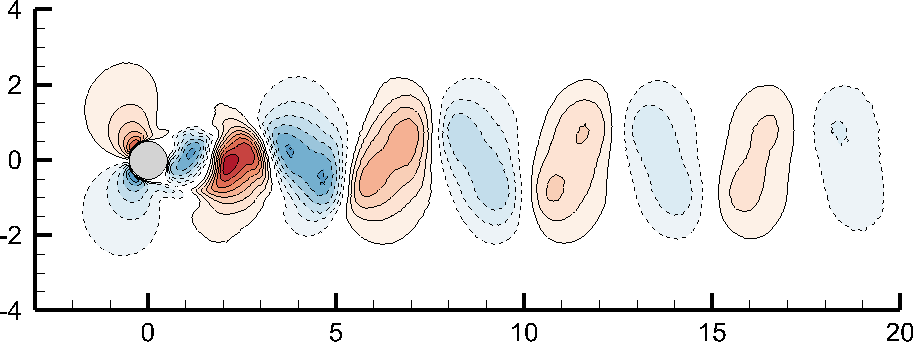}
\caption{Evolution of the transverse velocity at the $z=0$ plane; $Re=200$, 
$1/Fr=4$. The solid and dashed lines correspond to strictly positive and negative values, 
marked with contour interval $0.1V_o$. Going clockwise from the upper left panel, 
the corresponding dimensionless times are $T=30$, $62.5$, $125$, and $150$. \label{unsteady}}
\end{figure}

\begin{figure}
\centering
\includegraphics[width=0.8\linewidth]{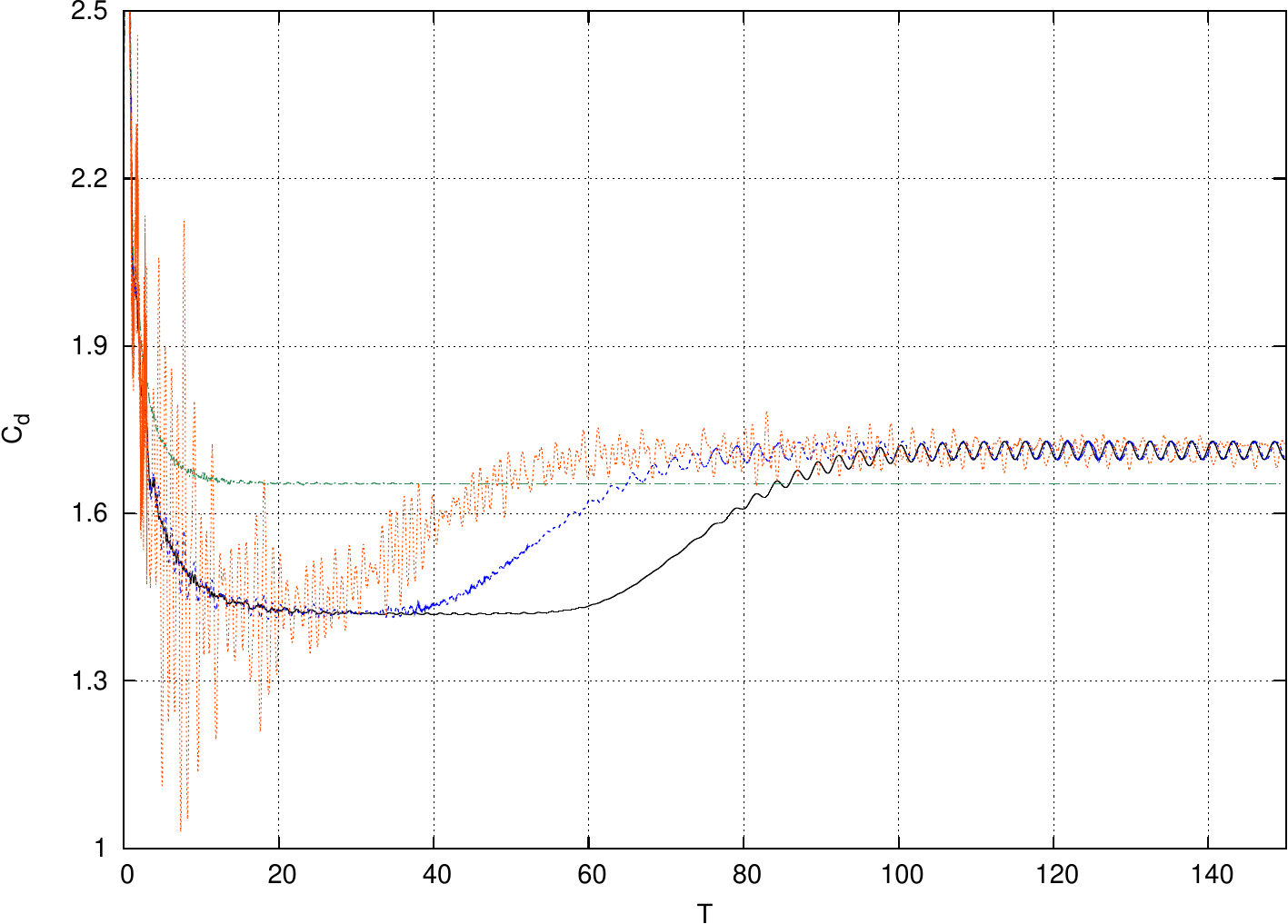}
\caption{Comparison of the drag coefficient histories for the reference 
setup (solid black) with analogous setups but with noisy initial conditions: 
noise amplitudes $0.2V_o$ (dashed blue) and $2V_o$ (dotted red), respectively, 
and the first-order-accurate upwind advection in lieu of MPDATA (dashed-dotted 
green); $Re=200$ and $1/Fr=4$. \label{cumdrag}}
\end{figure}

The computations for $1/Fr=4$ in \cite{szsm19} used a serial code and,
following \cite{hanazaki}, adopted the non-dimensional simulation time
$T=30$. The current results show that for stratifications with 
$1/Fr>2.5$, a longer simulation is required to trigger a transition from 
a steady to an unsteady pattern. The four panels in Fig.~\ref{unsteady} 
illustrate the evolution of the flow pattern for $1/Fr=4$ that gradually 
transitions from the symmetric steady state reported in \cite{hanazaki} to a
quasi-two-dimensional vortex shedding. In experimental and computational
studies, the transition to unsteady flow is typically triggered by arbitrary
perturbations. Indeed, the addition of a white noise with amplitude $0.2V_o$
introduced into the potential flow initialisation substantially accelerates
the transition, and the onset of shedding is sensitive to the magnitude
of the perturbations, Fig.~\ref{cumdrag}. However, after long simulation times, 
all computed flows developed the same vortex shedding pattern as that 
shown for $T=150$ in Fig.~\ref{unsteady}. Furthermore, a deliberate test 
replacing the second-order-accurate MPDATA in the NFT-FV solver with an 
over-diffusive first-order upwind scheme, shows that excessive numerical diffusion 
can entirely prevent the transition to an unsteady flow regime. Physically 
this is not surprising as the onset of shedding is related to the viscous 
boundary layer separation whereas the established flow depends robustly on
the Reynolds and Froude numbers. Numerically, however, these results attest
to the solution quality and robustness on the unstructured non-symmetric meshes.

\begin{figure}[h]
\centering 
\includegraphics[width=0.8\linewidth]{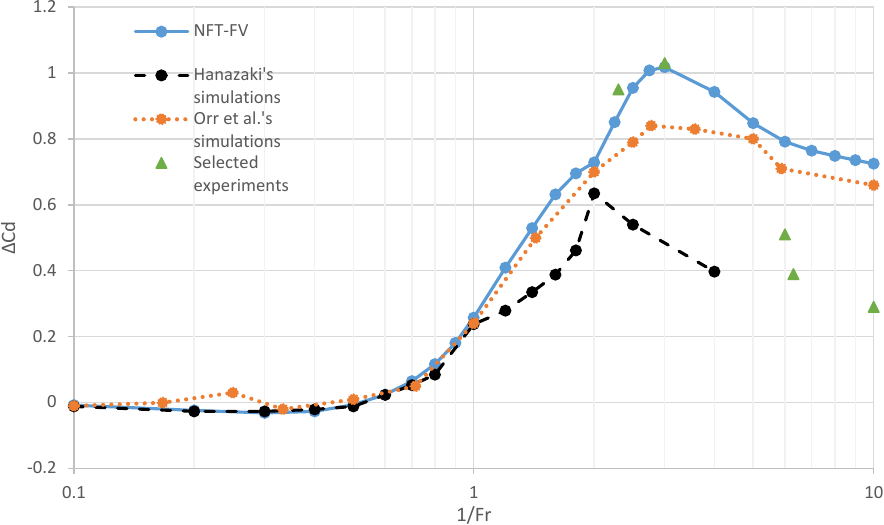}
\caption{$\Delta C_d$ in function of the inverse Froude number $1/Fr$
for numerical solutions at $Re=200$, NFT-FV (solid blue), \cite{hanazaki}
(dashed black), and \cite{orr} (dotted orange) as well as the experimental
data \cite{lofquist} (green triangles). \label{deltdrag200}} 
\end{figure}

Following \cite{hanazaki}, the influence of stratification on the
drag is quantified in terms of the departure of the drag coefficient 
$\Delta C_d=C_d(Re, 1/Fr)-C_d(Re, 0)$ from its value for the neutrally 
stratified flow. Figure~\ref{deltdrag200} compares our calculations for 
$1/Fr \in [0,10]$ with the results obtained experimentally
in \cite{lofquist}, and computationally in \cite{hanazaki} and
\cite{orr}. For unsteady flows the drag coefficient in our
calculations was averaged over the time interval $T \in [100,
150]$. Since the experimental data in \cite{lofquist} were obtained
at varying Reynolds numbers, Fig.~\ref{deltdrag200} shows their
results only from the experiments at $Re$ closest to $200$. For
clarity, these values are listed in Table~\ref{tbl1}.  
Up to $1/Fr=1$, all results are in good agreement. For $1/Fr\in[0.1,0.5]$, the
stratification is weak, so $\Delta C_d$ remains close to zero. For
$1/Fr>1$ the results of \cite{hanazaki} show a significant departure
from other data. For $1/Fr\in[1,2]$, the remaining computations
are in excellent agreement.  In the range $1/Fr\in[2,3]$,
the NFT-FV results match the experimental values closely and
show an increasing slope. At $1/Fr=3$, the results from experiment,
computation of \cite{orr} and the current simulation all show  a sharp
decrease in slope. This change of regime is consistent with the linear
theory of \cite{smith80,smith88} and the numerical/laboratory results
in \cite{smrot89,vosper}. Between $1/Fr\in[3.5 ,5]$, experimental data
is only available for different Reynolds numbers, however, both \cite{orr}
and the current NFT-FV computations show similar decreasing slope
tendencies as the experiments in \cite{lofquist}. For $1/Fr\in[5,10]$,
both computations give comparable results but their departure from the
experiments is significant. The authors in \cite{orr} attribute this
to an insufficient mesh resolution for a strong stratification which
could also be a factor affecting the NFT-FV results. We shall return
to this issue in section \ref{sec:Conclusion}.
\vspace{0.4\baselineskip}
\begin{table}[h]
\caption{$\Delta C_d$ values of \cite{lofquist} used in Fig.~\ref{deltdrag200}. \label{tbl1}}
\begin{tabular}{cccccccccccc}
\hline
\hline
 $1/Fr$ & $\vert$ & $2.3$ & $3$ & $6$ & $6.3$ & $10$ \\
\hline
$\Delta C_d$ & $\vert$ & $0.95$ & $1.03$  &  $0.51$ & $0.39$ & $0.29$\\
\hline
$Re$ & $\vert$ & $183$ & $221$ & $226$ & $215$ & $191$ \\
\hline
\end{tabular}
\end{table}

Further quantitative analysis compares the Strouhal numbers evaluated
from the periodic evolution of the drag and side force coefficients at
$1/Fr=3, 5, 10$. Table~\ref{tbl2} displays the NFT-FV results together
with the corresponding outcome obtained in \cite{lee04}. The results are
in good agreement overall, with the magnitude of $St$ evaluated from
the drag coefficient being typically twice as large as that reached by
$St$ computed from the side force coefficient. The latter is consistent
with an interpretation of the 2D vortex shedding from the sphere being
akin to the vortex shedding in a flow past a cylinder \cite{lee04}. The
Strouhal numbers evaluated from the lift coefficients (not shown)
obtained by the NFT-FV scheme are close to those obtained from the side
force coefficients.
\vspace{0.4\baselineskip}
\begin{table}[h]
\caption{Strouhal numbers based on $C_d$ and $C_s$ coefficients (cf. \cite{lee04}). \label{tbl2}}
\begin{tabular}{cccccccccc} 
\hline 
\hline 
$1/Fr$    & $\vert$ & 10~ (10) & 5~ (5) & 3~ (3.3) \\
\hline
$St(C_d)$ & $\vert$ & 0.395~ (N/A) & 0.376~ (0.398) & 0.395~ (N/A) \\
\hline
$St(C_s)$ & $\vert$ & 0.198~ (0.202) & 0.198~ (0.196) & 0.198~ (0.172) \\
\hline
\end{tabular}
\end{table}

\subsection{Flow past a sphere at $Re=300$}

As observed experimentally (Figs.~39a,b in \cite{johnson1999flow}) and
numerically (e.g. Fig.~2c in \cite{gus16}), the neutrally stratified
flow past a sphere at $Re=300$ results in periodic vortex shedding.
The drag coefficient obtained from our NFT-FV simulation for $Re=300$
and $1/Fr=0$ is evaluated by averaging the drag force over the
non-dimensional time interval between $T=250$ and $T=300$. Its mean
value $C_d=0.664$ and oscillation amplitude $C'_d=0.002$ are consistent
with the results in \cite{szsm19} and the references therein. The
Strouhal number derived from the frequency spectra of instantaneous
$C_d$ is $St=0.138$. It compares well with $St=0.136$ and $St=0.137$
reported in \cite{johnson1999flow,con03} and \cite{tomboulides00,yoon09},
respectively.

\begin{figure}[h]
\centering
\includegraphics[width=0.69\linewidth]{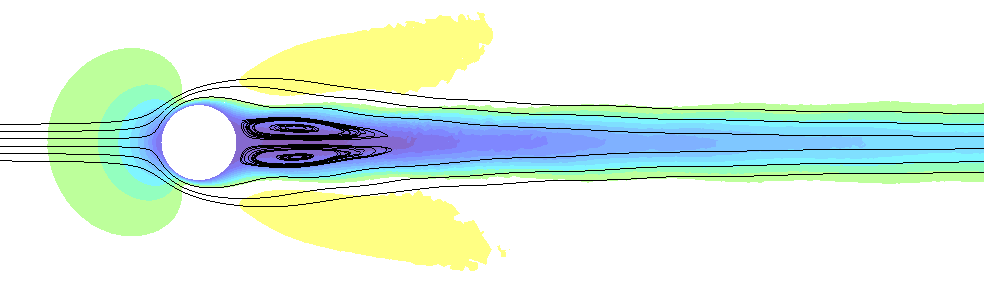}
\includegraphics[width=0.69\linewidth]{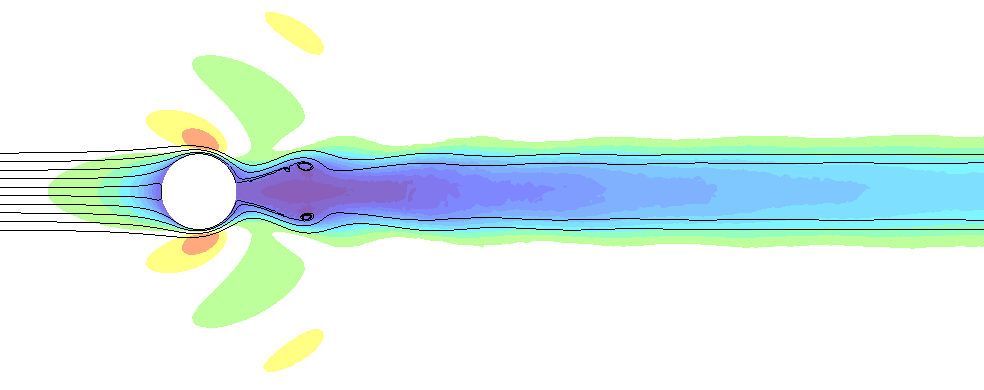}
\includegraphics[width=0.59\linewidth]{figures/horizvelo1lab.png}
\caption{Contours of streamwise velocity component and the associated 
streamlines plotted in the planes $z=0$ (top) and $y=0$ (bottom), 
respectively; $Re=300$ and $1/Fr=1.6$. \label{re300fr1.6} }
\end{figure}

For weak stratification, $1/Fr=0.1$, the character of the neutrally
stratified flow vortex pattern is maintained; see Fig.~\ref{wakes}
illustrating 3D features in the near wake and contrasting them with
a distinctly different $1/Fr=0.1$ flow at $Re=200$. The units of the
periodic wake structure consist of hairpin vortices with planar symmetry
preserved in time and set by the incipient shedding. Three vortex-shedding 
structures can fit in the adopted computational domain. As the
stratification increases the vortex shedding is suppressed and the flow
becomes close to steady-state, Fig.~\ref{re300fr1.6}, apart from tiny 
recirculation regions that form due to the combined effects of buoyancy 
and shear, just like for $Re=200$. Figure~\ref{wakes} shows that for
lower Froude numbers, $1/Fr=1.6$ and $4$, stratification dominates viscous
effects and the flow patterns for $Re=200$ and $Re=300$ become 
close. The results for Froude numbers where the lee-wave instability
regime occurs are also similar for both Reynolds numbers, with the
lee-waves pattern being already well-developed at $1/Fr=1.6$. However, for
$Re=300$ the change from the lee-waves instability into two-dimensional
vortex shedding structure occurs around $1/Fr=1.8$; i.e. earlier than
for $Re=200$. Furthermore, a closer inspection of Figs.~\ref{re300fr1.6}
and \ref{horivelo16a} shows that at the same $1/Fr=1.6$, the length of
the steady recirculation formed behind the sphere in the horizontal plane
increased by comparison with that measured for the $Re=200$ case.

\begin{figure}[h]
\centering \includegraphics[width=0.69\linewidth]{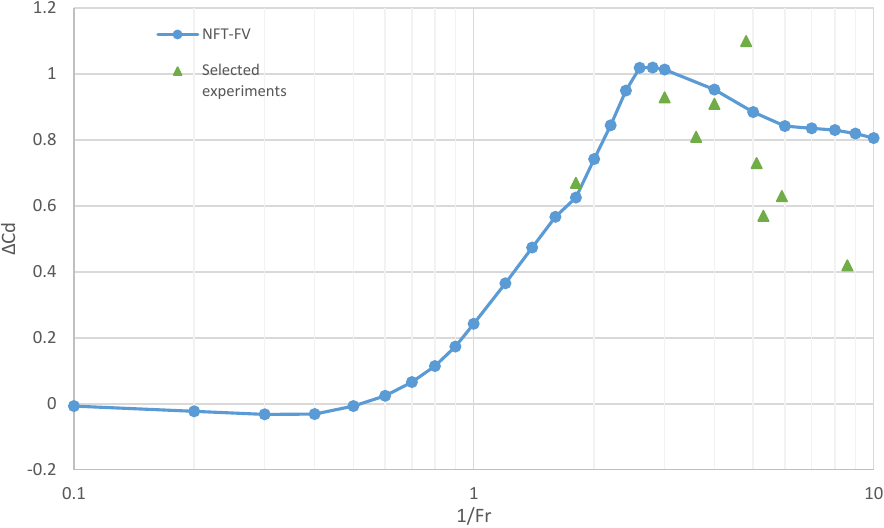}
\caption{$\Delta C_d$ in function of the inverse Froude number $1/Fr$
for numerical solutions at $Re=300$; NFT-FV calculations (blue line) and 
experimental data of \cite{lofquist} (green triangles). \label{deltdrag300} }
\end{figure}

Figure~\ref{deltdrag300} compares the NFT-FV results with the experiment
\cite{lofquist} for $\Delta C_d$ as a function of $1/Fr$. As for the case
of $Re=200$, the drag coefficient in the NFT-FV calculations was averaged 
over the time interval $T \in [100, 150]$. Only the data for the Reynolds 
numbers close to $300$, listed in Table~\ref{tbl3}, are shown in the graph. 
The results are well matched for $1/Fr=1.8$ and display appropriate slope
changes up to $1/Fr \sim 3$. The results for
$1/Fr=3$ and $1/Fr=4$ are close to the experimental values. Notably,
the experimental result for $1/Fr=3.6$ does not follow the slope
defined by experiments at $1/Fr=3$ and $1/Fr=4$. This and the
similar discrepancy for $1/Fr=4.8$ are possible manifestations of the magnitude of
experimental uncertainty. However when $1/Fr>5$, the decline of
the slope for the simulation results is systematically shallower than
in the experiment. A direct comparison of $\Delta C_d$ for $Re=200$
and $Re=300$ (not shown), revealed differences for $1/Fr \in [1.2,1.8]$
which could be associated with an earlier change to the two-dimensional
vortex shedding regime at $Re=300$. Differences are also present for
strongly stratified flows at $1/Fr \in [5,10]$. These are expected since
Fig.~\ref{wakes} indicates that the amplitude of lee waves for $Re=300$
in the vicinity of the sphere is larger than for $Re=200$. Furthermore,
a larger physical viscosity elicits a delayed development of a
periodic vortex shedding. For example, for $1/Fr=4$ the transition to
the fully developed periodic flow takes place at $T \approx 100$ for
$Re=200$ but occurs at $T \approx 60$ for $Re=300$.
\vspace{0.4\baselineskip}
\begin{table}[t]
\caption{$\Delta C_d$ values of \cite{lofquist} used in Fig.~\ref{deltdrag300}. \label{tbl3}}
\begin{tabular}{cccccccccccc}
\hline
\hline
 $1/Fr$ & $\vert$ & 1.8 & 3. & 3.6 & 4 & 4.8 & 5.1 & 5.3 & 5.9 & 8.6 \\
\hline
$\Delta C_d$ & $\vert$ & 0.67 & 0.93 & 0.81 & 0.91 & 1.1 & 0.73 & 0.57 & 0.63 & 0.42 \\
\hline
$Re$ & $\vert$ & 331 & 326 & 272 & 339 & 283 & 319 & 251 & 318 & 263 \\
\hline
\end{tabular}
\end{table}

\begin{figure}[!h]
\centering
\includegraphics[width=0.8\textwidth]{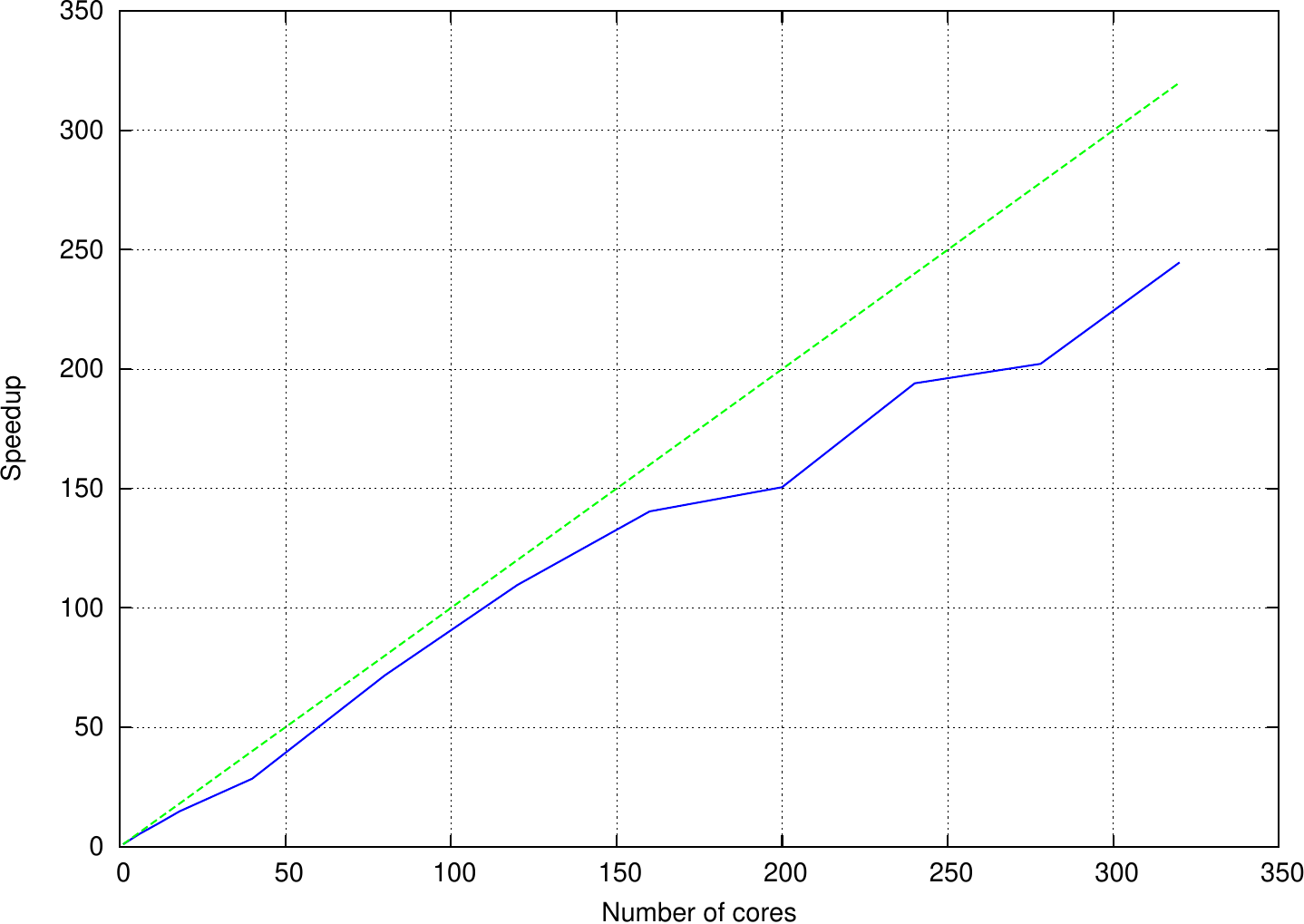}
\caption{Scaling results for the MPI parallelised NFT finite-volume code, 
for $Re=300$, $Fr=1$; dotted line shows ideal scaling.} 
\label{scaling}
\end{figure}

The new parallel implementation of the NFT-FV code operating on
fully unstructured meshes was extensively tested for the flow
past a sphere at $Re=300$. The computational mesh employed in the
presented calculations (described in \S{\ref{ssec:pformul}}) is highly
irregular, with the node spacing varying from $0.005D$ in proximity
to the sphere surface to $1.00D$ at the outer boundaries of the model
domain. Figure~\ref{scaling} highlights the code's speed-up, tested for
$Re=300$ and $Fr=1$. The speed-up ratio $T_{1}/T_{N}$, compares the
computational time of the sequential and the $N$-processes runs needed to
complete the simulation. The NFT-FV code shows good scaling until $160$
cores, where the parallel efficiency is $0.88$ on Hydra's Intel Xenon
CPU E5-2680v2 HPC.\footnote{http://hpc-support.lboro.ac.uk/} Beyond that,
the parallel performance becomes affected by the limitations of the mesh
size and partitioning variables of the MeTis library.

\subsection{Flow past two spheres} 
\begin{figure}[h] 
\centering
\includegraphics[width=0.6\linewidth]{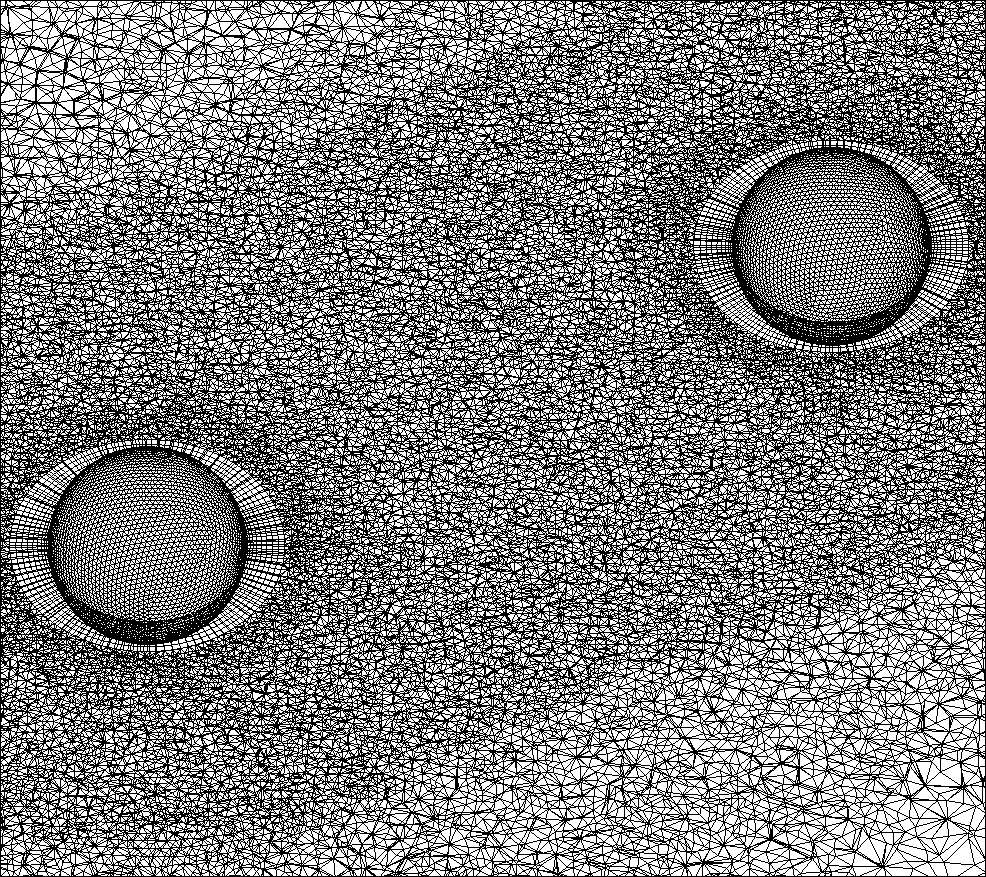}
\caption{A fragment of the $y=0$ cross-section of the primary mesh for 
a flow past two spheres and the triangular surface mesh. \label{mesh} }
\end{figure} 

To illustrate flexibility of unstructured meshes, the NFT-FV
solver was applied to simulate flows past two spheres at $Re=300$. Two
identical, $D=1$, spheres were placed in a tandem arrangement aligned
with a free stream flow. The coordinates of the first sphere's centre
are $(0,0,0)$, while the centre of the second sphere is located
at $(4,0,0)$. To allow the flow to fully develop downstream of the
second sphere, the computational domain is extended to form a cuboid
$[-15,29]\times[-15,15]\times[-15,15]$. The hybrid computational mesh,
shown in Fig.~\ref{mesh}, is built in the same way as for the single
sphere domain, with 10 triangular base prismatic layers added around
each sphere. The mesh consists of $1927331$ nodes which are more densely
concentrated around and inbetween the two spheres and in the wake regions. The smallest
distance of $0.005D$ between points corresponds to the thickness of
the first prismatic layer from the sphere surfaces, while the largest
elements of the primary mesh reach the approximate dimensions of $1D$
near the outer boundaries. This configuration and setup have been chosen
to facilitate a quantitative comparison with the neutrally stratified
flow solution at $Re=300$ in \cite{yoon09}.

\begin{figure}[h]
\centering
\includegraphics[width=0.8\linewidth]{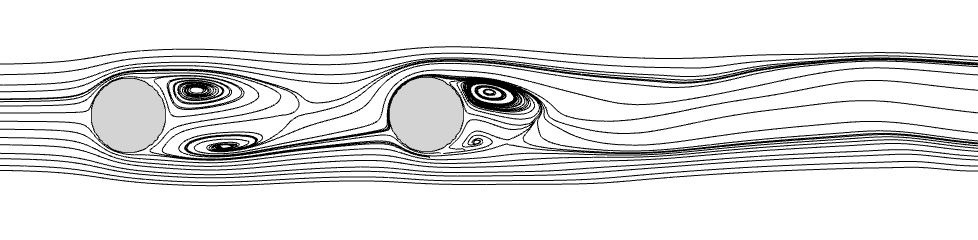}
\caption{Streamline plot in the central vertical plane for the $1/Fr=0$
flow past two spheres at $Re=300$. Streamwise direction extent is
$x\in [-1.7, 11.5]$. \label{fig:nonstrat_ts-y}}
\end{figure}

The neutrally stratified flow past the two spheres was simulated
for $T=400$. Figure~\ref{fig:nonstrat_ts-y} shows the streamline
plot in the vertical central plane, $y=0$. The streamlines reveal
non-symmetric recirculations in the lee of both spheres. The shear
layer generated by the windward sphere becomes reattached to the
second sphere and interacts with the second sphere's boundary layer.
This affects the wake behind the second sphere and decreases the
size of the largest recirculation by a factor of $0.58$ in comparison
with the analogous single sphere case. 

\begin{figure}[h] 
\centering
\includegraphics[width=0.99\linewidth]{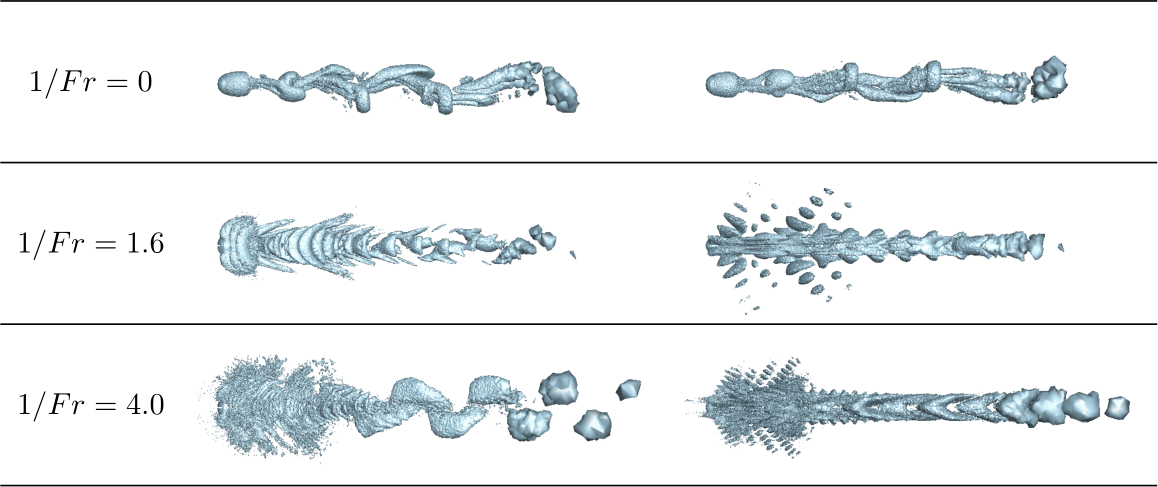}
\caption{Flow past the tandem spheres at $Re=300$ and $1/Fr=0, 1.6, 4$; wake vortex 
structures in the horizontal (left) and vertical (right) central planes. \label{fig:nonstrat_ts}}
\end{figure}

The upper row of
Fig.~\ref{fig:nonstrat_ts} displays the 3D flow pattern obtained
for $Re=300$ and $1/Fr=0$. The $Q$-method visualisation shows that
the hairpin vortices characteristic of the Reynolds number $300$
flow past a single sphere (cf. \cite{szsm19}) are also developing
in the present case. Furthermore, the right panel clearly shows
how the wake generated by the first sphere overlaps the second 
sphere. Similarly, as with the single sphere in \cite{szsm19},
the symmetry plane for the tandem configuration for the neutrally
stratified flow is formed in the diagonal plane, rotated $45^{\circ}$
about the $X$-axis. Notably, even a weak stratification imposes
the vertical position of the symmetry plane; cf. $Re=300, 1/Fr=0.1$
case in Fig.~\ref{wakes}.

\begin{figure}[h] 
\centering
\includegraphics[width=0.49\linewidth]{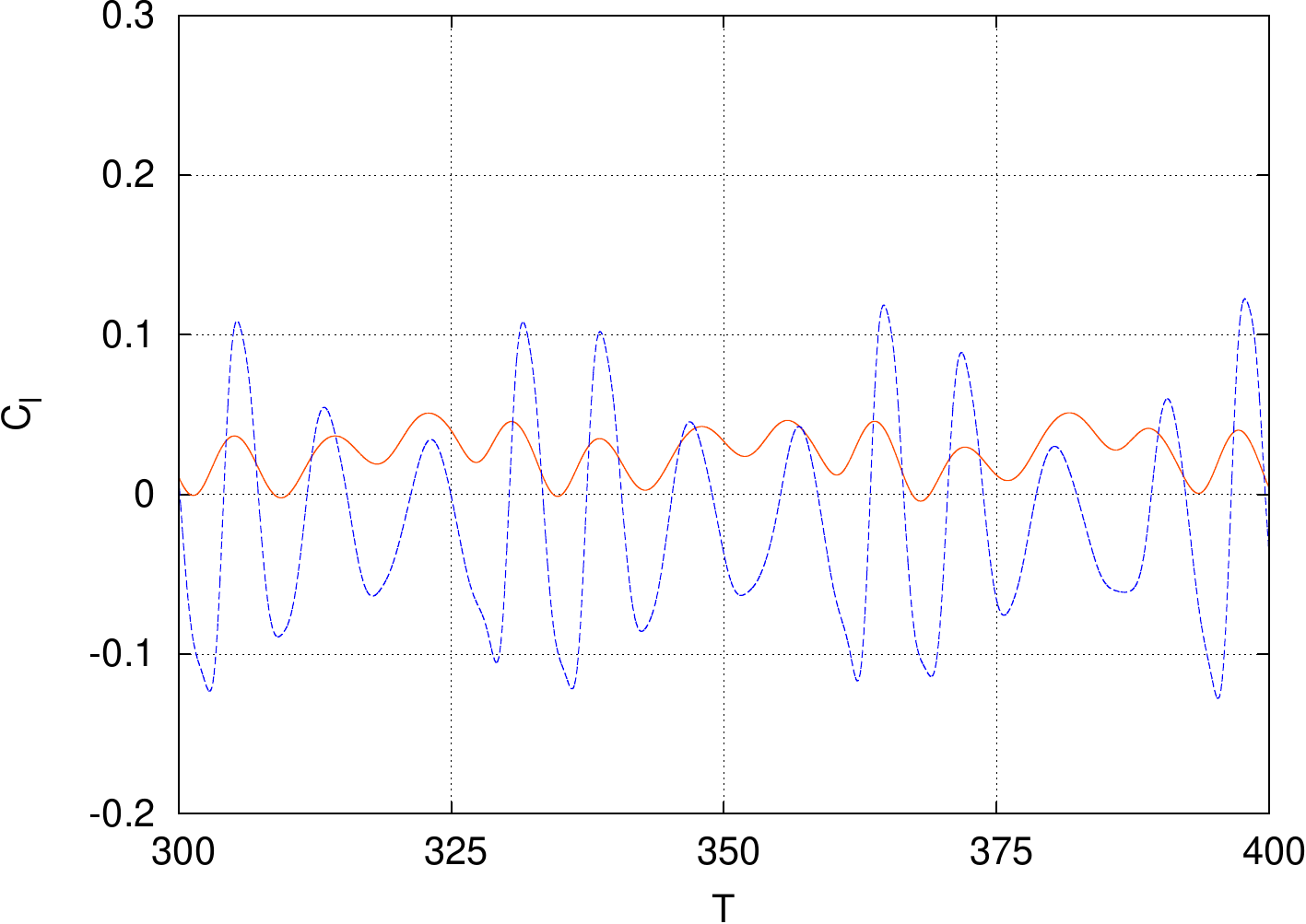}
\includegraphics[width=0.49\linewidth]{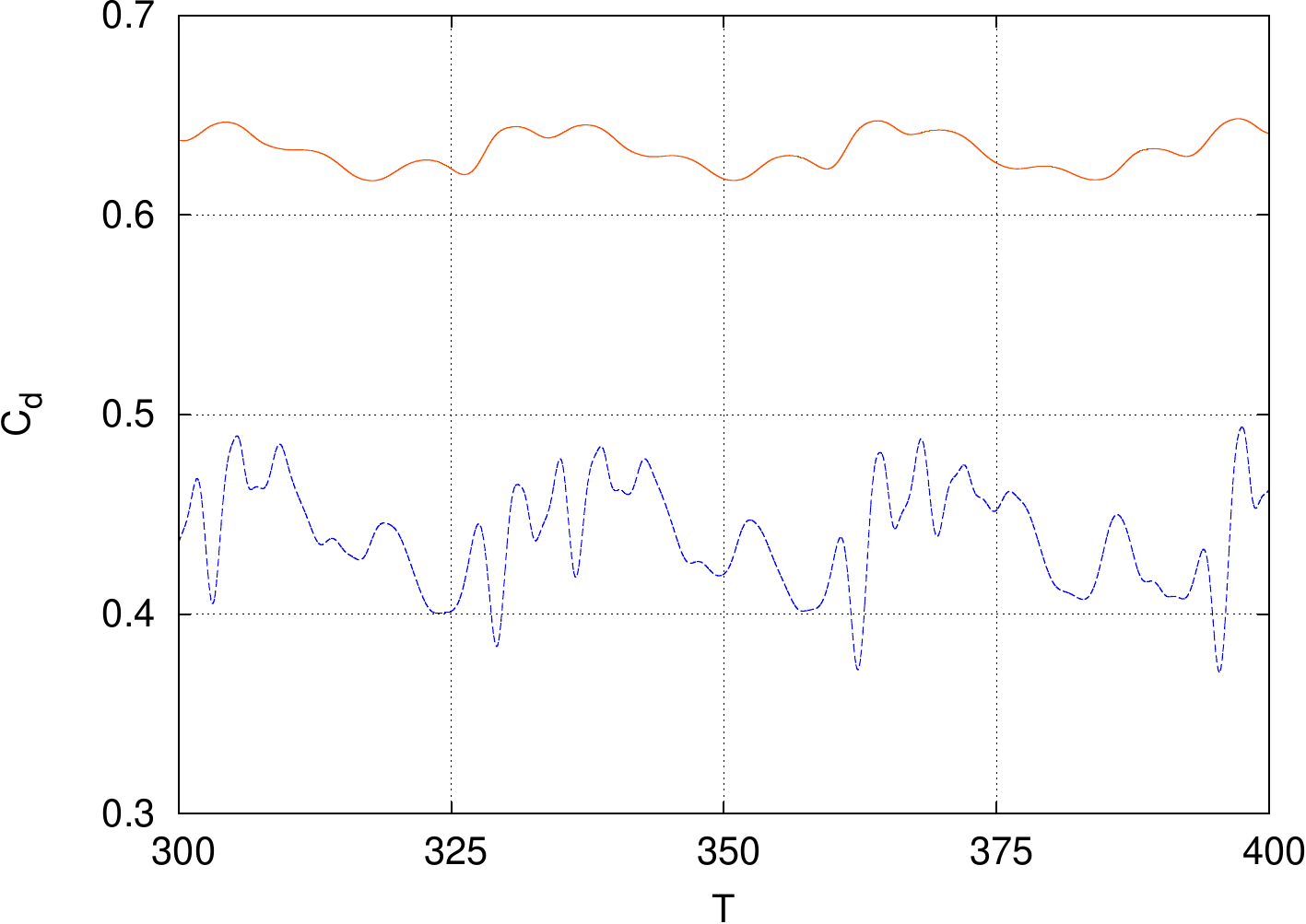}
\caption{History of the lift (left) and drag (right) coefficients 
in $Re=300$ and $1/Fr=0$ flow past two spheres; solid (red) and 
dashed (blue) lines correspond to the first and second sphere values.}
\label{freq300}
\end{figure}

Multiple vortex-shedding frequencies for each sphere were reported in
\cite{yoon09}. The left panel of Fig.~\ref{freq300} shows the history
of the lift coefficient obtained for both spheres. Frequencies from
the NFT-FV simulation match closely those of Fig.~16 in \cite{yoon09}.
The phase is shifted due to different initial conditions and the amplitude
differs slightly due to the rotation of the symmetry plane in our
calculations. The right panel complements the left panel with the history
of the drag coefficient. The corresponding mean drag coefficients are
$C_d=0.632$ and $C_d=0.439$ for the first and second sphere respectively.


Two dominant vortex-shedding frequencies $f_1$ and $f_2$ are found
for each sphere. Table~\ref{tab:freq} provides the Strouhal numbers
$St_1=f_1 D / V_o$ and $St_2=f_2 D / V_o$, where $f_1$ and $f_2$
are derived from the frequency spectra of both drag and lift
coefficients (not shown) in the time interval $T\in [300, 400]$.
They are in a good agreement with those reported in \cite{yoon09}.
Overall, these results reflect the influence of the first sphere's
leeward flow reattachment in generating the vortices behind the
second sphere.
\vspace{0.4\baselineskip}
\begin{table}[h] \caption{Strouhal numbers for the $Re=300$, 
$1/Fr=0$ flow past two spheres.}
\begin{tabular}{ccccc} \hline \hline
 & \multicolumn{2}{c}{First sphere} & \multicolumn{2}{c}{Second sphere} \\
 & $St_1$ & $St_2$ & $St_1$ & $St_2$ \\
\hline
  NFT-FV($C_d$) & 0.119 & 0.030 & 0.089 &0.030 \\ 
  NFT-FV($C_l$) & 0.119 & 0.030 & 0.119 & 0.149 \\
  \cite{yoon09}($C_l$) & 0.122 & 0.030 & 0.122 & 0.153 \\
 \hline
\end{tabular}
\label{tab:freq}
\bigskip
\end{table}

\begin{figure}[t]
\centering
\includegraphics[width=0.8\linewidth]{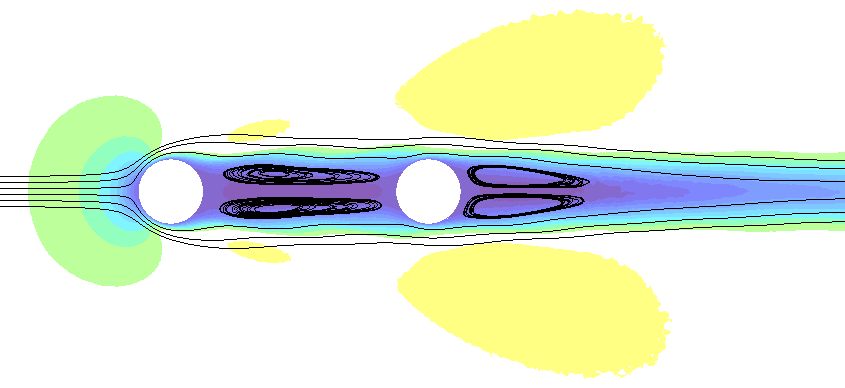}
\includegraphics[width=0.8\linewidth]{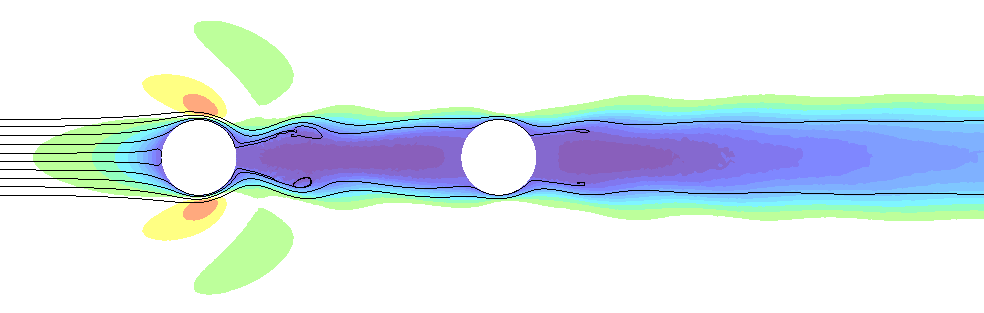}
\includegraphics[width=0.59\linewidth]{figures/horizvelo1lab.png}
\caption{Contours of streamwise velocity component and the associated 
streamlines plotted in the planes $z=0$ (top) and $y=0$ (bottom), 
respectively; $Re=300$ and $1/Fr=1.6$. \label{2sphk16} } 
\end{figure}

The flow past two spheres is dramatically influenced by the strength 
of stratification, as was previously observed for the single sphere. For
$1/Fr=1.6$, the symmetric steady state flow is established demonstrating
the features characteristic to the lee-wave instability regime behind both
spheres, Fig.~\ref{2sphk16}. Consistent with the single sphere
case, the vertical scales are inhibited by the stratification. Additionally, 
in comparison to Fig.~\ref{re300fr1.6} the length of steady recirculation 
formed behind the first sphere in the horizontal plane is approximately 
one diameter longer and stretches in the entire gap between the spheres.
The recirculation after the second sphere is smaller and closer in length 
to the single sphere recirculation length. The drag coefficients computed
for the first and the second sphere are $C_d=1.100$ and $C_d=0.531$. 

Boundary layer separates behind both spheres, as shown in the lower panel of Fig.~\ref{2sphk16} for a vertical cross-section at $y=0$.
The separation angles, denoted by $\phi_s$, are measured clockwise from the upstream centerline on the vertical plane, where $\phi_s=164\degree$ and $\phi_s=145\degree$ for the first and second sphere respectively.
In comparison, the separation angle for the single sphere at the same Reynolds and Froude numbers (Fig.~\ref{re300fr1.6}) is $\phi_s=166\degree$, demonstrating that the separation angles behind the first sphere and the single sphere are similar.
The first sphere blocks the incoming flow slowing down its motion between the spheres, and the flow separation limits the width of the wake induced by the first sphere. These two factors lead to a substantially reduced separation angle for the second sphere. 

Similarly, for the strongly stratified flow at $1/Fr=4$, the separation angle for the first sphere, $\phi_s=131\degree$, is higher than for the second sphere, $\phi_s=121\degree$, while the value for first sphere is lower than $\phi_s=138\degree$ observed for the corresponding flow past a single sphere.
The middle and bottom panels of Fig.~\ref{fig:nonstrat_ts} show 3D plots
for flows with $1/Fr=1.6$ and $4$. The two-dimensional vortex-shedding
regime at $1/Fr=4$ also preserves the main features of the flow
past a single sphere. As was observed for $1/Fr=0$ and $1.6$, the mean drag
coefficients for $1/Fr=4$ evaluated for the first sphere, $C_d=1.477$,
is close to that obtained for the single sphere case.  However, the
mean drag coefficient computed for the second sphere, $C_d=0.960$,
is significantly lower, reflecting the first sphere's blocking of the
flow seen by the second sphere.  The corresponding Strouhal numbers
are $St_1(C_d)=0.020$ and $St_2(C_d)=0.327$ for the first sphere,
and $St_1(C_d)=0.020$ and $St_2(C_d)=0.337$ for the second sphere.
The Strouhal number for the second sphere is similar to that for the
single sphere, indicating similarities in the vortex shedding in the
far wake, visible also in Fig.~\ref{fig:nonstrat_ts}. In contrast,
the Strouhal number for the first sphere is strongly affected by the
presence of the obstacle in its immediate wake.

\section{Concluding remarks \label{sec:Conclusion}}

A study exploring behaviour of stratified flows past spheres have been
conducted using a newly developed parallel option of the nonoscillatory
forward-in-time scheme operating on flexible unstructured meshes and
employing finite volume spatial discretisation. The study repeated well
documented benchmarks for neutrally-stratified flows at $Re=200$ and $300$,
allowing for validation of the proposed numerical approach. Additionally,
for the stably stratified flows at $Re=200$, earlier numerical and
experimental studies provided further means for qualitative and
quantitative validation of the NFT-FV model.

Our study of stably stratified flows at $Re=300$ provides further
insights into stratified flows at moderate Reynolds numbers.
Detailed documentation of numerical simulations and their analysis
was supplemented with comparisons with experimental data when available.
The presented results confirmed the general experimental characterisation
for stratified flows past a single sphere, introduced in \cite{lin92}.
In particular, they showed that the overall flow patterns are dominated
by stratification as the Froude number decreases. Although entirely
different at high Froude numbers---where the flow at $Re=200$ reaches
the steady state and at $Re=300$ results in a periodic hairpin vortex
shedding---at $1/Fr=1.6$ both flows become planar and are in the lee-wave
instability regime. In this regime, both flows reach steady state except 
for tiny recirculations in the lee indicating the lee-waves breaking. 
Differences between the two flows are in the details, such as the lengths of the
recirculations formed behind a sphere that can be observed in the central
horizontal plane. This similarity between the flow patterns continues 
at lower Froude numbers where the flows enter a two-dimensional vortex
shedding regime in the horizontal, while remaining planar in the vertical,
e.g.\ at $1/Fr=4$. Details of the stably stratified flow features, such as 
amplitudes and length of gravity waves, slightly differ between the two 
Reynolds numbers. For $Re=200$, we also document the triggering of a transition
to the 2D vortex shedding regime at $1/Fr=2.25$. The exact transition
between the regimes is dictated by viscosity and occurs at lower Froude 
number for lower Reynolds numbers. 

In addition, the single sphere study was extended to akin simulations
of flows past two spheres, facilitated in the framework
of unstructured meshes. The results for the neutrally stratified flow at
$Re=300$ are in good agreement with those reported in the literature \cite{yoon09}
and further confirm the validity of the NFT-FV model. The stratified
flow past two spheres is, to our knowledge, documented for the first
time and investigated for the chosen tandem configuration at $Re=300$
with $1/Fr=1.6$ and $1/Fr=4$. The resulting flow patterns generally
follow that recorded for the single sphere, however, a detailed analysis
reveals particularities of the spheres' interactions that can be quite
complex. These interactions can be associated with intricacies
of atmospheric/oceanic flows past mountain ranges, islands and complex
bathymetry \cite{LeeMAP87,Reisner94,WarnVOM07}, since they depend on 
the relative placement of the two spheres \cite{yoon09}. The tandem sphere
simulations illustrate the potential of the NFT-FV solver operating on flexible
meshes to provide a further insight into the class of stratified flows
pertinent to engineering and geophysics, suggesting interactions between
multiple spheres. 
Our preliminary simulations indicate that, similarly as for
neutrally stratified flows, the flow patterns for multiple spheres
are sensitive to the distance between the spheres and their relative
location with respect to the flow direction. Identification of the
flow regimes for configurations of multiple spheres will be studied
in our future work.

Regardless of the overall satisfactory outcome of the study, it remains
puzzling why numerical results for flows past a single sphere depart
from experimental measurements for strong stratifications with $1/Fr>5$,
Figs.~\ref{deltdrag200} and \ref{deltdrag300}. Furthermore, the results
using two distinct numerical models---ours and that of \cite{orr},
available for $Re=200$---give drag coefficients that agree well with each
other, Fig.~\ref{deltdrag200}. The numerical results using yet
another model \cite{gushchin06}, available for $1/Fr=5$ and $Re=100$,
also appear to overestimate the drag coefficient. The design of numerical
experiments is essentially the same for our and \cite{orr} calculations,
where reduction of the Froude number is achieved by increasing the
ambient stratification $S$ (defined in \S{\ref{ssec:pdes}}). However
in \cite{gushchin06} the stratification is fixed, and the Froude number
reduction is achieved by increasing the sphere's radius. The authors in
\cite{orr} attribute the drag discrepancy to the loss of the resolution,
which is plausible because as $Fr\searrow 0$ the amplitude of the fluid
parcel vertical displacements, $\eta$, scales as $\eta=0.5 D Fr$,
based on the dividing streamline arguments \cite{szsm19}. However,
based on the same arguments, increasing the sphere radius effectively
fixes the magnitude of the displacements and thus their effective
resolution. Consequently, the result of \cite{gushchin06} is at odds
with the insufficient resolution hypothesis. Moreover, as the only common
element in all three numerical designs is the Boussinesq approximation
adopted in the governing equations, one might speculate invalidity of
the Boussinesq model in the limit of small Froude numbers. Indeed, in
our and \cite{orr} experiments the ratio of the parcel displacements and
the density (or potential temperature) height scale $S^{-1}$ increases as
$Fr\searrow 0$. Yet this ratio $\eta S$ is fixed in \cite{gushchin06},
which is again at odds with such speculations. Running out of ideas on
how to blame the numerical results, we turn to experimental data. The
design of the laboratory experiments in \cite{lofquist} was special, in
that the size of the sphere and the ambient stratification were fixed with
the buoyancy frequency $N\propto Re/Fr$ (cf. \S{5.2} in \cite{szsm19} for
discussion), effectively associating the Froude number reduction with the
threading (viz.  free stream) velocity. In such a design $Re\searrow 0$
as $Fr\searrow 0$, with $C_d$ and $\Delta C_d$ vanishing in the limit. On
one hand, this shows an elementary incompatibility of these experimental
data with the numerical experiments discussed, or vice versa. On the
other hand it adds to the intricacy of the $Fr\searrow 0$ limit, potentially
rejuvenating interests in the asymptotic \cite{drazin} theories.

\bigskip
\noindent {\it Acknowledgements:}
This  work  was  supported in part by the funding received from the
EPSRC studentship grant 1965773, Horizon 2020 Research and
Innovation Programme (ESIWACE Grant agreement no. 675191, ESCAPE
Grant agreement no. 671627 and ESCAPE2 Grant agreement no. 800897).
This publication reflects the views only of the authors, and the
Commission cannot be held responsible for any use which may be made of
the information contained therein. The computation in this paper used
the 'Hydra' High Performance System at Loughborough University. NCAR is
sponsored by the National Science Foundation.

\end{document}